\documentclass[PRD,onecolumn,showpacs,preprintnumbers,nofootinbib,amsmath,amssymb]{revtex4}

\usepackage{amssymb,color}
\usepackage{graphicx}
\usepackage{dcolumn}
\usepackage{bm}

\def\be{\begin{equation}}
\def\ee{\end{equation}}

\begin{document}

\title{Gamma Rays from Clusters and Groups of Galaxies: Cosmic Rays versus Dark Matter}

\author{Tesla E.~Jeltema$^1$\footnote{Morrison Fellow}, John Kehayias$^2$ and Stefano Profumo$^{2,3}$}
\affiliation{$^1$ UCO/Lick Observatories, Santa Cruz, CA 95064, USA\\
$^2$Department of Physics, University of California, Santa Cruz, CA 95064, USA\\
$^3$ Santa Cruz Institute for Particle Physics, University of California, Santa Cruz, CA 95064, USA
}

\date{\today}

\begin{abstract}
\noindent Clusters of galaxies have not yet been detected at gamma-ray frequencies; however, the recently launched Fermi Gamma-ray Space Telescope, formerly known as GLAST, could provide the first detections in the near future. Clusters are expected to emit gamma rays as a result of (1) a population of high-energy cosmic rays fueled by accretion, merger shocks, active galactic nuclei and supernovae, and (2) particle dark matter annihilation. In this paper, we ask the question of whether the Fermi telescope will be able to discriminate between the two emission processes. We present data-driven predictions for the gamma-ray emission from cosmic rays and dark matter for a large X-ray flux limited sample of galaxy clusters and groups. We point out that the gamma-ray signals from cosmic rays and dark matter can be comparable. In particular, we find that poor clusters and groups are the systems predicted to have the highest dark matter to cosmic ray emission ratio at gamma-ray energies. Based on detailed Fermi simulations, we study observational handles that might enable us to distinguish the two emission mechanisms, including the gamma-ray spectra, the spatial distribution of the signal and the associated multi-wavelength emissions. We also propose optimal hardness ratios, which will help to understand the nature of the gamma-ray emission. Our study indicates that gamma rays from dark matter annihilation with a high particle mass can be distinguished from a cosmic ray spectrum even for fairly faint sources. Discriminating a cosmic ray spectrum from a light dark matter particle will be instead much more difficult, and will require long observations and/or a bright source. While the gamma-ray emission from our simulated clusters is extended, determining the spatial distribution with Fermi will be a challenging task requiring an optimal control of the backgrounds.
\end{abstract}


\maketitle


\section{Introduction}

Clusters and groups of galaxies are the largest gravitationally bound matter structures observed in the Universe. Although these objects are expected to host several high energy phenomena, the resulting electromagnetic non-thermal emission is far from being fully understood \cite{brunetti,heclusters}. A hallmark of the occurrence of non-thermal phenomena in these large structures is the detection, in numerous clusters, of extended radio emission associated to the synchrotron losses of relativistic cosmic-ray electrons \cite{brunetti,heclusters,Berrington:2002bh,Colafrancesco:1998us,Reimer:2004au}. The acceleration of cosmic rays in galaxy clusters can originate from a number of physical processes, including violent shocks produced in cluster-cluster mergers and the accretion of smaller structures \cite{harris,sarazin,shocks,Keshet:2002sw}, the re-acceleration of cosmic rays injected by galactic sources like active galactic nuclei and supernovae \cite{sarazin02}, and inelastic collisions of primary cosmic-ray protons  \cite{dennison} producing showers of secondary particles, including relativistic electrons and positrons as well as gamma rays. 

The radio emission from clusters, perhaps the most solid source of observational information on non-thermal phenomena in these  objects, broadly falls into two classes featuring different spatial distribution, polarization and emission location within the cluster. The first class,  {\em radio relics}, is characterized by irregular radio morphologies and is typically located in external regions of the cluster \cite{kempner}. The second class of diffuse cluster radio sources is that of {\em radio halos}  \cite{giovannini}, whose emission is typically centered on the cluster and follows a similar spatial distribution as e.g. the thermal X-ray emission from the intra-cluster medium (ICM) gas. Radio relics, unlike radio halos, exhibit prominent polarization, and are often associated to cluster regions thought to host shock activity. The origin of radio halos, instead, is far from being fully understood \cite{brunetti,heclusters}. The upcoming generation of low-frequency radio arrays, including the Giant Metrewave Radio Telescope (GMRT), the LOw Frequency ARray for radio astronomy (LOFAR), the Murchison Widefield Array (MWA) and the Long Wavelength Array (LWA), will improve the current observational situation in the near future.

In the recent past, claims of hard X-ray emission from nearby cluster of galaxies have been reported \cite{hardxray}. One possibility is that this X-ray emission originates from the inverse Compton (IC) scattering off background radiation of the same non-thermal high energy electron population responsible for the radio emission. Most of the clusters reportedly detected at hard X-ray frequencies are merging clusters \cite{nevalainen2004}, and the detections themselves are still debated \cite{rossetti04,Ajello:2008rd}. In addition, it is unclear if simple models for the non-thermal population fueling radio and hard X-ray emissions can self-consistently explain data from clusters (see e.g. the case of new radio data from the Ophiuchus cluster of galaxies \cite{ophradio} in connection with the recent claim of a non-thermal hard X-ray detection with INTEGRAL \cite{ophhxr}). Future hard X-ray missions, including the Nuclear Spectroscopic Telescope Array (NuSTAR) \cite{nustar} and the International X-ray Observatory (IMX) \cite{syx}, will play a key role in settling the mentioned controversies, as well as in fostering our understanding of processes underlying non-thermal activity in clusters \cite{Ajello:2008rd}.

Gamma rays, covering the highest end of the electro-magnetic spectrum, can potentially provide essential information on high-energy phenomena in groups and clusters of galaxies. So far, these objects have not been detected in gamma-rays, and data from the EGRET telescope on board the Compton Gamma-Ray Observatory \cite{egretref} have only produced upper limits \cite{reimer} (see however \cite{grdetect}). Statistically conclusive discoveries and a true revolution in our understanding of the highest energy phenomena in galaxy clusters are anticipated with the Fermi Gamma-ray Space Telescope, formerly known as GLAST \cite{glastref}, which was successfully launched on June 11, 2008. The main instrument onboard Fermi, the Large Area Telescope (LAT), represents a remarkable jump in sensitivity compared to its predecessor EGRET \cite{egretref}. The LAT also extends the EGRET energy range (20 MeV to 10 GeV) to much higher gamma-ray energies, up to about 300 GeV. Aside from probing high and ultra-high energy particle physics processes in astrophysical sources, the detection of gamma rays from clusters would also help to establish the properties of the primary proton population within clusters, and possibly clarify its role in various cluster phenomena \cite{heclusters}. These include e.g. the question of the origin of radio halos \cite{dennison}, the ``cooling-flow'' problem \cite{fabian} and particle acceleration within cluster merger shocks \cite{sarazin04}.

Clusters of galaxies are potentially powerful observational probes of cosmology (see e.g. \cite{voit,HMH01, albrecht06}). In this context, the accurate understanding of non-thermal phenomena in clusters is crucial to their ultimate utility as cosmological probes. Specifically, the estimate of cluster masses, which is at the basis of all cosmological applications using clusters, frequently relies on the assumption of hydrostatic equilibrium between gravitational forces and the thermal pressure supplied by the ICM. The accuracy of hydrostatic mass determinations is therefore limited by our understanding of the non-thermal pressure provided by cosmic rays, turbulence and magnetic fields in the ICM \cite{ensslin97,nagai2007}. The Fermi Gamma-Ray Space Telescope, as well as current and planned ground based atmospheric Cherenkov telescopes (ACTs) will be able to probe the energy density supplied by cosmic rays, and possibly the evolution with redshift of cosmic ray pressure, to an unprecedented level of accuracy \cite{pfrommer2004,nagaiando}.

A more exotic possibility for non-thermal activity in galaxy clusters was first envisioned by Totani in Ref.~\cite{totani}: the pair annihilation of weakly interacting massive particles (WIMP) constituting the dark matter halo. Following that seminal work, Colafrancesco, Profumo and Ullio calculated in \cite{Colafrancesco:2005ji} the complete multi-frequency spectrum, for the case of the Coma cluster, resulting from dark matter annihilation. The emission spectrum extends from radio to gamma-ray frequencies, and includes the secondary emissions from the non-thermal electrons and positrons produced as final stable particles in dark matter annihilation events. In addition, Ref.~ \cite{Colafrancesco:2005ji} also studied the heating of the ICM produced by the energy injected by dark matter annihilation, as well as the induced Sunyaev-Zeldovich signal. As far as indirect signals from particle dark matter annihilation, by far the best studied non-thermal radiative emission is the production of gamma-rays \cite{Gunn:1978gr,Stecker:1978du}. Specifically, when two dark matter particles annihilate, gamma rays result both directly from loop-suppressed diagrams as well as from the subsequent decays or radiative emission (e.g. from final state radiation)  of standard model particles produced in the annihilation final state, like quarks, leptons and gauge and Higgs bosons. The resulting gamma rays have energies extending up to the kinematic limit set by the WIMP mass (the pair annihilation event occurs for highly non-relativistic dark matter particles), predicted to be in the 10-1000 GeV range in the best motivated models \cite{dmreviews,kkdm} (see however \cite{profumoheavy,profumolight}). 

The LAT instrument onboard Fermi is a tremendous tool for the indirect search for particle dark matter with gamma rays \cite{Baltz:2008wd,ourgc}. Of special relevance for the present study, the flux of gamma rays from WIMP dark matter annihilation in clusters of galaxies has been shown to be in principle large enough to be detectable by Fermi-LAT \cite{Colafrancesco:2005ji,ophiuchus}. Dark matter annihilation does not only produce gamma rays, but also additional stable particle species, such as energetic electrons and positrons. These, in turn, produce synchrotron, IC and bremsstrahlung radiation, with unique spectral features. The multi-wavelength emission from dark matter annihilation was studied in detail in \cite{gondolo,bertone0101134,aloisio0402588,baltzwai,coladraco,xrdwarf,ullioregis,haze1,haze2}, and specifically in clusters of galaxies in \cite{Colafrancesco:2005ji,ophiuchus,colabullet}. Interestingly, it was demonstrated that if dark matter annihilation fuels to some appreciable degree either the radio emission in the Coma cluster \cite{Colafrancesco:2005ji}, or the hard X-ray emission in the Ophiuchus cluster \cite{ophiuchus}, Fermi is almost guaranteed to have the sensitivity to detect gamma rays from dark matter annihilation.

Upcoming gamma ray observations of clusters of galaxies have therefore profound implications for cosmology and, possibly, the discovery of New Physics. If Fermi detects gamma rays from clusters, a crucial point will be to conclusively assess the nature of the mechanism responsible for the emission. In the present study, we focus on how to tell apart gamma rays produced by standard, astrophysical mechanisms such as cosmic rays from those resulting from WIMP dark matter annihilation. On general grounds, we expect to have three handles to differentiate the two emission mechanisms:
\begin{itemize}
\item the {\em gamma-ray spectrum}: models for cosmic ray production of gamma rays predict a flux as a function of energy which differs from what expected out of dark matter annihilation. The low photon statistics represents a challenge for meaningful discrimination based on the gamma-ray spectrum. We thus study the best angular and energy range, and propose a technique based on hardness ratios that will help to discriminate cosmic rays from dark matter
\item the {\em spatial distribution}: depending upon assumptions on the dark matter substructure distribution and density profile, as well as on the primary cosmic ray source distribution, we predict that clusters can appear to be extended gamma-ray sources. We study whether this can be used to differentiate gamma rays emitted by dark matter from those produced by cosmic rays 
\item the {\em multi-wavelength emission}: comparing the results of hydrodynamical simulations of cosmic rays in clusters to our predictions for a dark matter scenario, we find that the ratio of the hard X-ray to gamma-ray emission is a potential diagnostic to understand the origin of non-thermal phenomena in clusters.
\end{itemize}
In our simulations and analysis, we use the latest LAT instrumental response function and observation strategy and the Fermi Science Tools software package which is currently being used to analyze Fermi data.

As an application of our theoretical study, we present predictions for gamma-ray fluxes from a large X-ray limited sample of 130 nearby groups and clusters of galaxies. Specifically, our predictions are based on X-ray data, and on a fixed set of assumptions for both dark matter and cosmic rays. This also allows for a meaningful comparison of the dark matter to cosmic ray induced emission at gamma-ray frequencies. We present a ranking of plausible candidates where one might expect a bright gamma-ray signal, and of sources where the dark matter contribution is expected to be stronger compared to the one fueled by astrophysical cosmic rays. In particular, we discovered that low-redshift groups are the most promising class of objects to search for a dark matter signal from distant extra-galactic systems. 

The organization of our paper is as follows. The following Section \ref{sec:method} illustrates the model we use to compute the gamma-ray emission resulting frtom cosmic rays (\ref{sec:cr}) and from dark matter annihilation (\ref{sec:dm}), and gives details on the Fermi simulation setup (\ref{sec:glastsim}). Section \ref{sec:origin} discusses how to study the origin of gamma rays from clusters, including our analysis of the optimal angular region, the spectra, hardness ratios, spatial extension and multi-wavelength counterparts. We present in Section \ref{sec:cat} our predictions and ranking of nearby clusters and groups according to their gamma-ray emission (the complete list is provided in the Appendix). Section \ref{sec:concl} gives a discussion and summary of our results, and concludes.

\section{Methodology}\label{sec:method}
In this Section we present our modeling of the gamma ray emission from cosmic rays (\ref{sec:cr}) and from dark matter (\ref{sec:dm}). For definiteness, we consider the case of the Coma cluster \cite{Colafrancesco:2005ji,grcoma}, but what we find applies to generic low redshift clusters. We also discuss, in Sec.~\ref{sec:glastsim}, the Fermi simulation setup we employ in our study.

\subsection{Gamma-ray Emission From Cosmic Rays}\label{sec:cr}

Several mechanisms leading to the acceleration of relativistic particles in the intra-cluster medium have been discussed in the literature (see e.g.~\cite{Reimer:2004ac}). Most importantly, energetic arguments suggest that powerful shocks created in cluster-cluster mergers and in the accretion of material onto the deep cluster gravitational potential well are significant sources of relativistic cosmic rays \cite{Takizawa:2000qk,shocks,Keshet:2002sw,Blasi:2003xs,Berrington:2002bh,Miniati:2001ay,2008arXiv0806.1522S,Colafrancesco:1998us}. The same shocks can also re-accelerate originally lower-energy particles injected into the ICM through other processes \cite{Ensslin:1997kw}.

The common denominator to the above mentioned scenarios is that Fermi shock acceleration yields a population of non-thermal relativistic cosmic rays. These include primarily high-energy electrons and protons. The former efficiently loose energy by synchrotron emission at radio frequencies as well as through the up-scattering of background radiation to gamma-ray and X-ray frequencies (inverse Compton scattering). Collisions of high-energy cosmic-ray protons with nuclei in the ICM produce, in their hadronic debris, neutral pions promptly decaying into two gamma rays with typical energies and fluxes potentially observable by Fermi-LAT. In addition, the same collisions yield secondary cosmic ray electrons and positrons from the decays of charged pions \cite{Berrington:2002bh,Petrosian:2001ph,Berezinsky:1996wx}.

At energies $E_\gamma\gtrsim 0.1$ GeV and for non-merging clusters, or clusters in the intermediate or late merger stages, most of the gamma-ray emission is believed to stem from gamma rays produced in neutral pion decays resulting fomr the above mentioned inelastic cosmic-ray proton collisions \cite{Berrington:2002bh}. The energy stored in the secondary electron-positron pairs is predicted to contribute at the level of 1\% or less of the total power associated to the primary cosmic ray protons \cite{Berrington:2002bh}.  In terms of the gamma-ray emission, ref.~\cite{nagaiando} estimates that only for very low average cluster magnetic fields and for very steep proton injection spectral indexes can the secondary inverse Compton contribution be  even 10\% of that from $\pi^0$ decay. 

Although secondary $e^\pm$ inverse Compton emission is likely subdominant compared to neutral pion decay gamma-ray yields, during the early stages of a merger, {\em primary} electrons can still make a significant contribution to the GeV radiation.  The gamma-ray emission is, again, dominantly associated to the inverse Compton of, here, primary cosmic-ray electrons off of the microwave radiation background. The resulting inverse Compton flux is suppressed compared to the hadronic gamma-ray production from nuclear interactions involving non-thermal protons only as long as the efficiency of acceleration of hadronic species exceeds that for electrons \cite{Keshet:2002sw}. While this is indeed the expectation in diffusive shock acceleration theory \cite{Berrington:2002bh}, the reader should bear in mind that the model we outline below might not apply to merging clusters in the early stage of a merger, or it may give underestimates of the cosmic ray production of gamma-rays. In general, inverse Compton emission from primary electrons is expected to dominate close to acceleration sites such as large scale shocks \cite{Miniati2002}. Nevertheless, simulations indicate that the inverse Compton emission from electrons is systematically subdominant at energies relevant to the Fermi telescope compared to gamma-rays produced from pion decays \cite{pfrommer2004, pfrommer2007,nagaiando}. 

In this work, we consider only the typically dominant gamma-ray emission from pion decays. The inclusion of gamma rays from primary electrons inverse Compton scattering could affect the results of the analysis we present in the following ways. First, the total gamma-ray flux, when considering this additional source, will generically be enhanced, potentially resulting in even better prospects for the detection of galaxy clusters at gamma-ray energies with Fermi-LAT. Furthermore, if IC from primary electrons plays a significant role, the morphology of the gamma-ray emission region from cosmic rays would also be affected, leading likely to a wider extent of the emitting region (including for instance peripheral cluster shock regions). This could, however, potentially hinder the discrimination of a cosmic-ray emission from that originating from dark matter annihilation using spatial considerations (sec.~\ref{sec:space}). Thirdly, the spectral analysis (sec~\ref{sec:specs}) will in general be affected, depending on the spectrum of the emission, and a broken power law feature could arise at low Fermi energies where the emission from primary electrons may become comparable to that from primary hadronic cosmic rays. Finally, it is possible that clusters that host a bright active nucleus  and, concurrently, exhibit significant IC emission from an accretion shock would feature a {\em double peak} in the spatial distribution of high-energy cosmic-ray sources. Such a circumstance cannot be described by the radial power-law functional form we adopt in the simplified spatial model for the distribution of cosmic-rays described below.

Bearing the above caveats in mind, for simplicity we assume here that the dominant source for the gamma-ray emission from galaxy clusters at energies relevant to Fermi originates from inelastic collisions of hadronic cosmic rays \cite{Reimer:2004ac,Berezinsky:1996wx}. Following the arguments outlined above, we then follow the analytical cosmic-ray model outlined in \cite{pfrommer2004, pfrommer2007, nagaiando}. In this scenario, the primary proton injection spectrum is described by a simple power law, parametrized by a spectral index $\alpha_p$ independent of the position in the source. In \cite{pfrommer2004} a framework is outlined for incorporating the fireball model for very high energy cosmic ray proton interactions with the ICM as well as pion production threshold effects.  The resulting differential source function $q_\gamma$ (with units of inverse energy and volume) is
\begin{eqnarray}\label{eq:crspec}
q_\gamma(r,E_\gamma) \textrm{d}E_\gamma\textrm{d}V \simeq& \sigma_{pp}\  c\  n_N(r)\ \xi^{2-\alpha_p}\ \frac{\tilde{n}_{CR_p}(r)}{\textrm{GeV}}\ \frac{4}{3\alpha_p}\left(\frac{m_{\pi^0}c^2}{\textrm{GeV}}\right)^{-\alpha_p} \nonumber \\ &\times \left[\left(\frac{2E_\gamma}{m_{\pi^0}c^2}\right)^{\delta_\gamma}+\left(\frac{2E_\gamma}{m_{\pi^0}c^2}\right)^{-\delta_\gamma}\right]^{-\alpha_\gamma/\delta_\gamma}\ \  \textrm{d}E_\gamma\textrm{d}V,
\end{eqnarray}
where 
\be
\sigma_{pp} = 32\left(0.96+e^{4.4-2.4\alpha_p}\right)\textrm{mbarn}
\ee
models the effective inelastic p-p cross section and $n_N$ is the target nucleon density in the ICM. The quantity $\tilde{n}_{CR_p}$ with the dimensions of the cosmic ray proton number density, has a normalization chosen so that the kinetic cosmic ray proton energy is proportional to the thermal energy density of the ICM. Also, $\xi = 2$ is the pion multiplicity, and $\delta_\gamma = 0.14\alpha_p^{-1.6}+0.44$ is a shape parameter for the $\pi^0$-threshold physics \cite{pfrommer2004}.  By integrating over all solid angles and dividing by $n_N$ and $\tilde{n_{CR_p}}$ (to be independent of a model's spatial dependence) the final differential gamma-ray source function (i.e. the gamma-ray flux per unit energy and unit time, per unit target and impinging cosmic-ray flux) is obtained.

The spatial distribution of cosmic ray sources is modeled in terms of the ratio of the energy density in cosmic rays to the energy density of the thermal gas; this ratio is taken to be a power law with radius, parametrized by $\beta_p$ as
\begin{equation}\label{eq:spatial}
X_p(r) = X_p(R_{500})\left(\frac{r}{R_{500}}\right)^{\beta_p}
\end{equation}
where $X_p$ is the ratio of the energy density of cosmic rays compared to the thermal gas, and $R_{500}$ is the radius of an enclosed spherical overdensity $500$ times the critical density of the universe at the source's redshift \cite{nagaiando}. The parameter $\beta_p$ physically reflects the possibility that the spatial distribution of the sources of high-energy cosmic rays deviates, via a power-law as a function of radius, from the density profile of the thermal gas in the cluster.  While a variation in the Mach number during the course of a cluster merger can drive different injection indexes at different locations \cite{Berrington:2002bh}, we neglect here, for simplicity, any spatial variation associated to $\alpha_p$.

In summary, the simple model we use here depends on three orthogonal parameters: $\alpha_p$, that sets the spectral shape of both the injected primary cosmic rays and the resulting gamma rays, $X_p$ that sets the normalization of the gamma-ray flux, and $\beta_p$, that (together with the gas density distribution, that can be inferred e.g. from X-ray data) sets the spatial distribution of the signal. We outline below a few motivated parameter space choices which define the benchmark models we employ to run our Fermi simulations. We then apply these models to the specific case of the Coma cluster. 

\begin{itemize}
\item As far as the injection spectral index, estimates come from theoretical arguments as well as from numerical simulations. For instance, Ref.~\cite{Berrington:2002bh} showed that the minimum spectral index ranges from 2.1 to 2.8. In that range, larger values are typically associated to forward shocks and smaller masses. The larger the mass, the stronger the gravitational potential and the harder the predicted injection spectral index. Structure formation shock theory predicts injection spectral indices of $2.0<\alpha_p<2.5$ for strong shocks \cite{Miniati:2003ep}, such as those expected for {\em accretion} shocks and {\em strong merger} shocks. 
We choose here $\alpha_p=2.1,\ 2.7$ as physically motivated cases.  Specifically, the cosmic ray spectrum in our Galaxy is observed to be a power law with $\alpha_p=2.7$, which motivates the large $\alpha_p$ choice.  On the other hand, clusters confine cosmic rays on cosmological time scales \cite{Volk:1994zz,Berezinsky:1996wx}, and are thus expected to give rise to a harder spectrum than that of the Galaxy: $\alpha_p=2.1$ is thus also a reasonable and motivated choice for a harder injection spectrum \cite{nagaiando}. This choice is, in addition, consistent with the results of the simulations of Ref.~\cite{Berrington:2002bh}. However, if cosmic rays in galaxy clusters are accelerated in {\em weaker} merger shocks the expectation is one of a softer injection spectral index. 
In this case, strong confinement of cosmic rays in clusters will likely not make the spectrum much harder. Numerical simulations, though, indicate a low efficiency for the acceleration of cosmic rays at weak merger shocks \cite{kangjones}. In summary, thus, we regard our benchmark value $\alpha_p=2.7$ as an example case rather than, strictly, as an upper limit to the cosmic ray injection spectrum.

\begin{table}
\begin{tabular}{ccc}
\hline
Model & $\alpha_p$ & $\beta_p$\\
\hline
CR\_HC & $2.1$ & $-0.5$\\
CR\_HF & $2.1$ & $1$\\
CR\_SC & $2.7$ & $-0.5$\\
CR\_SF & $2.7$ & $1$\\
\hline
\end{tabular}
\caption{Summary of parameters for cosmic ray models used. The parameter $\alpha_p$ indicates the primary cosmic-ray injection spectral index, while the coefficient $\beta_p$ stands for the bias of the cosmic-ray source spatial distribution with respect to the cluster's thermal gas spatial distribution (see eq.~\ref{eq:spatial}). For all models, we set the ratio of the energy density of cosmic rays compared to the thermal gas $X_p=0.1$. \label{tab:cr_models}}
\end{table}

\item The results of recent hydrodynamical simulations (see e.g.~\cite{pfrommer2007}) and of other recent studies \cite{nagaiando} motivate our choices of $\beta_p=-0.5,1$ for our benchmark models.  The case of $\beta_p=-0.5$, which we label as the ``cuspy'' profile, is found in simulations that include radiative effects. The profile in (\ref{eq:spatial}) with $\beta_p=-0.5$ approximates the profile resulting from the radiative, hydrodynamical simulations of Ref.~\cite{pfrommer2007}.  Non-radiative simulations indicate instead a flatter profile with $\beta_p=1$; this is an extreme scenario where the density of cosmic ray sources relative to the thermal gas density grows linearly with radius. Although this might be on the verge of being unrealistic, the effects of cooling and heating in clusters are also somewhat uncertain, and we adopt this case as an extreme possibility \cite{pfrommer2007,nagaiando}. We remark that in recent studies \cite{brunetticassano} it has been argued that cosmic ray activity in clusters is dominated by turbulent reacceleration. In this case, in the absence of shocks, the natural expectation for $\beta_p$ would be 0. This choice falls in between the two benchmark cases we consider here.

\item Lastly, in our Fermi simulations we assume that the cosmic ray energy density is 10\% of the thermal gas energy density, i.e. we set $X_p=X_p(R_{500})=0.1$. The choice of the normalization to the cosmic ray proton energy density is limited by a few constraints \cite{Reimer:2004ac}: 
\begin{itemize}
\item The gamma rays produced in inelastic collisions yielding neutral pions, which subsequently decay into two photons, must be consistent with the EGRET upper limits \cite{reimer}
\item The inverse Compton up-scattering of microwave photons by the secondary $e^{\pm}$ population produced by charged pion decay (as for the neutral pions these particles are produced in collisions of the primary cosmic ray protons with nuclei in the ICM) will give rise to non-thermal hard X-ray and soft gamma-ray emission. The intensity of this radiation must also be consistent with observational data.
\item Lastly, the mentioned secondary $e^\pm$ will also radiate at radio frequencies via synchrotron emission, providing additional constraints from radio data.
\end{itemize}
Assuming $\alpha_p\simeq2.4$, Ref.~\cite{Reimer:2004ac} shows that in the case of Coma the radio data are at the level predicted for $X_p\sim0.2$, with appropriate assumptions on the magnetic field distribution. Tighter constraints come from high frequency radio data, and depend quite sensitively on $\alpha_p$ and, more importantly, on the value of the average magnetic field. EGRET data, again for the case of Coma, put milder constraints ($X_p\lesssim0.45$ for $\alpha_p=2.1$ and $X_p\lesssim0.25$ for $\alpha_p=2.5$ \cite{pfrommer2004}). While we consider $X_p=0.1$ here in our Fermi simulated observations, we will consider a more conservative value of $X_p=0.01$ in our survey of galaxy clusters and groups in sec.~\ref{sec:cat} and in the Appendix, motivated e.g. by the results of Ref.~\cite{nagaiando}, which showed that Fermi could be sensitive (depending on $\alpha_p$) to $X_p$ smaller than a fraction of a percent for nearby massive clusters.

\end{itemize}

We summarize our benchmark cosmic ray models, with their names and parameters, in Table \ref{tab:cr_models}. ``CR'' stands for cosmic rays, while ``H'' and ``S'' respectively indicate a ``\textit{Hard}'' and a ``\textit{Soft}'' primary proton injection spectrum, $\alpha_p=2.1$ and 2.7, and ``C'' and ``F'' stand for ``\textit{Cuspy}'' and ``\textit{Flat}'', respectively corresponding to $\beta_p=-0.5$ and to 1.

\begin{figure}[t]
\begin{center}
\includegraphics[width=14.cm,clip]{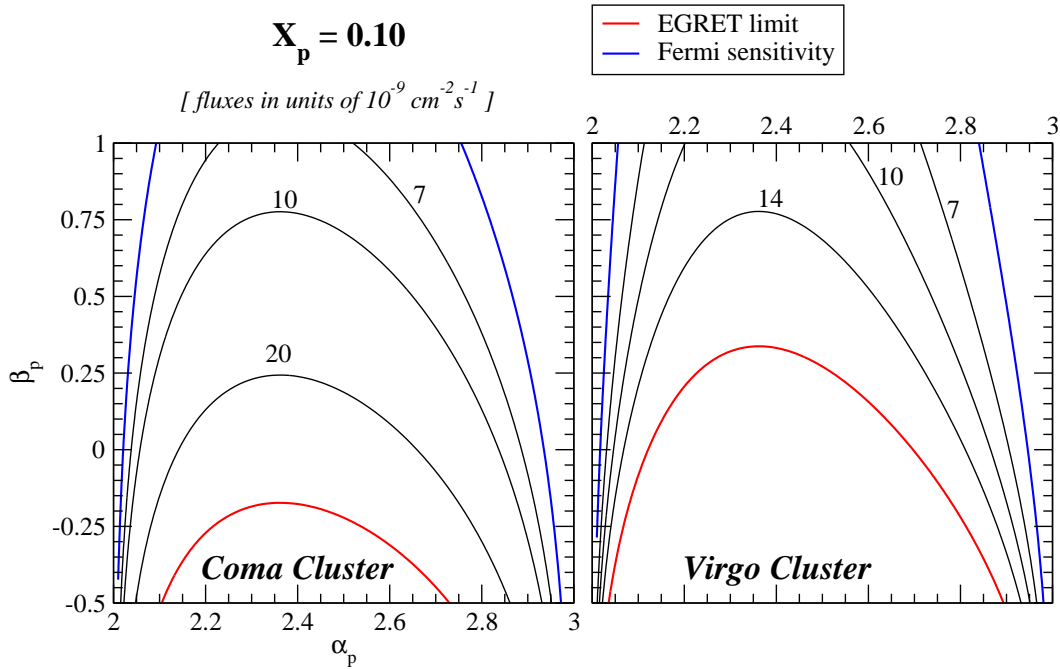}\\
\caption{Contours levels for the gamma-ray emission from cosmic rays in the Coma (left) and Virgo (right) clusters of galaxies. We assume that the cosmic rays have 10\% of the gas energy density ($X_p=0.1$), and scan the plane $(\beta_p,\alpha_p)$ where $\beta_p$ characterizes the spatial distribution of cosmic ray sources, and $\alpha_p$ the primary proton injection spectrum. \label{fig:alphabeta}}
\end{center}
\end{figure}

As pointed out in previous studies, for a wide range of parameters Fermi will be able to detect a gamma-ray signal produced by cosmic rays: Figure \ref{fig:alphabeta} shows the gamma-ray flux for the Coma and Virgo clusters for a range of $\alpha_p$ and $\beta_p$, with $X_p=0.1$. In the computation, we assumed for both clusters the gas density profiles as given in Table 1 of \cite{pfrommer2004}. The curves indicate contours of constant integrated gamma-ray flux above 0.1 GeV, in units of $10^{-9}\ {\rm cm}^{-2}\ {\rm s}^{-1}$. For simplicity, we assume here a Fermi sensitivity for a point-source emission of $4\times10^{-9}\ {\rm cm}^{-2}\ {\rm s}^{-1}$, while the EGRET gamma-ray limits for Coma and Virgo are given in \cite{reimer}. Points below the blue line, in the central part of the panels, are predicted to be within the sensitivity of Fermi. The red lines indicate instead the EGRET limits: parameter space points below those lines are thus ruled out by current data \cite{reimer}. 

The shape of the gamma-ray emission contours is not unexpected: larger values of $\beta_p$ imply smaller gamma-ray fluxes, simply because they feature a flatter cosmic ray spatial distribution (see Eq.~\ref{eq:spatial}), which in turn integrates to smaller values. In addition, we find that the source function in Eq.~(\ref{eq:crspec}) implies that the largest gamma-ray fluxes correspond to intermediate values (2.1$\lesssim \alpha_p \lesssim2.7$) of the injection spectrum. 


\subsection{Dark Matter Models}\label{sec:dm}
The determination of the gamma-ray flux from dark matter annihilation in clusters depends on both the dark matter density distribution in the specific object under consideration and on assumptions on the dark matter particle model. For definiteness, we again consider here the case of the Coma cluster, for which a detailed study of the dark matter density distribution was carried out in \cite{Colafrancesco:2005ji}.

As far as the dark matter density distribution is concerned, we assume a Navarro, Frenk and White (NFW) profile \cite{nfw} for both the smooth component (with a scale radius $a$, to be defined in the Equations below) and for the radial distribution of substructure (with a biased scaling parameter $a^\prime\simeq7a$, as inferred from numerical simulations of e.g. \cite{Nagai:2004ac, Diemand:2004kx}\footnote{Notice that (i) a smaller $a^\prime/a$ ratio is inferred from a comparison of the galaxy radial density profile in clusters \cite{Diemand:2004kx} (see also \cite{DeLucia:2003xe}), and (ii) this ratio might vary from cluster to cluster (for instance, Ref.~\cite{Diemand:2004kx} finds that the ratio scales proportionally with the cluster concentration). We evaluated the uncertainty on the total annihilation signal for the setup we outline in this section stemming from considering the generous range $2<a^\prime/a<10$ and we find a variation of at most $\pm 10\%$ in the flux within one degree and of less than $4\%$ for the emission within 0.1 degrees, for the case of the Coma cluster.}. We adopt here the semi-analytic approach outlined in \cite{Colafrancesco:2005ji} to evaluate the contribution from the smooth host halo and from substructures. The relevant quantity for the computation of the dark matter annihilation signal is a number density of particle dark matter pairs, defined as:
\begin{equation} \label{eq:npair}
{\mathcal N}_{\rm pairs}(r) =\frac{\rho_{\rm m}^2}{2m^2_{\rm WIMP}}\Big(\frac{\left[\rho^\prime\  g(r/a)-f_s\ \tilde\rho_s\  g(r/a^\prime)\right]^2}{\rho_{\rm m}^2}+f_s\ \Delta^2 \frac{\tilde\rho_s\  g(r/a^\prime)}{\rho_{\rm m}}\Big).
\end{equation}
In the Equation above, the first line represents the contribution from the smooth part of the dark matter halo, while the second line encompasses the contribution from substructures. In particular, $\rho_{\rm m}$ indicates the present day mean matter density in the Universe, and for the function $g(y)$, as alluded to above, we assume the NFW prescription, i.e.
\begin{equation}\label{eq:nfw}
g(y)=\frac{1}{x\left(1+x\right)^2}.
\end{equation}
The normalization parameter $\rho^\prime$ and the scale radius $a$ can be expressed, for a given profile, as functions of the virial mass $M_{\rm vir}$ and of the virial concentration parameter $c_{\rm vir}$ (by ``virial'' we mean assuming an overdensity $\Delta_{\rm vir}\simeq343$, see the discussion in \cite{Colafrancesco:2005ji}). Following \cite{Colafrancesco:2005ji}, we take here for the Coma cluster a virial mass $M_{\rm vir}\simeq 0.9\times 10^{15}M_\odot h^{-1}$ and a concentration $c_{\rm vir}\simeq10$. The distance to Coma is set to 95 Mpc \cite{Colafrancesco:2005ji}.  Further, in Eq.~(\ref{eq:npair}) above we defined a reference substructure normalization parameter
\be
\tilde\rho_s\equiv\frac{M_{\rm vir}}{4\pi(a^\prime)^3\int^{R_{\rm vir}/a^\prime}{\rm d}y \ y^2\ g(y)}.
\ee
Finally, the substructure model is specified by the two parameters $f_s$ and $\Delta^2$. The first stands for the ratio of the total mass in subhalos over the total virial mass,
\be
\int_{M_{\rm cut}}^{M_{\rm vir}}{\rm d}M_s\ \frac{{\rm d}n}{{\rm d}M_s}M_s=f_sM_{\rm vir},
\ee
where ${{\rm d}n}/{{\rm d}M_s}$ indicates the sub-halo mass function, and $M_{\rm cut}$ the small scale cut off mass in the matter power spectrum \cite{Profumo:2006bv,othersubs}.  The second term, $\Delta^2$, indicates the weighed enhancement in the number density of dark matter particle pairs due to subhalos. For the definition of $\Delta^2$ and an extensive discussion on how to assess its value we refer the reader to \cite{Colafrancesco:2005ji}, which we follow here. $\Delta^2$ crucially depends upon the ratio between the concentration parameter in subhalos over that in isolated halos at equal mass. Given a structure formation model and a dark matter density profile, $\Delta^2$ and the mentioned ratio can be traded for each other. Notice that as Eq.~(\ref{eq:npair}) shows, $\Delta^2$ {\em is not} the usually quoted substructure boost factor. 

We employ here two sets of $f_s$ and $\Delta^2$, representing a very conservative setup with a suppressed contribution from substructure (``{\em Smooth}'', or ``S'' case) and one where instead substructures play a very significant role in setting the dark matter annihilation gamma-ray signal (``{\em Boosted}'', or ``B'' case). For the Smooth case, we assume that only 20\% of the mass is in substructures ($f_s$=0.2) and that the average concentration ratio of same mass host and sub-halos equals 2,  following what quoted in \cite{bullock}. The latter assumptions yields $\Delta^2\simeq 7\times 10^5$. In the Boosted setup, we instead assume $f_s$=0.5 and a concentration ratio of 4, implying $\Delta^2\simeq 7\times 10^6$. While smaller dark matter substructure are in principle possible, our present choices are realistic and compatible with the results of N-body simulations.

Another possibility for the overall cluster dark matter density distribution is one where the innermost profile is flat. This case can be physically motivated e.g. in the context of scenarios where angular momentum is effectively transfered between baryonic and dark matter in the process of baryon infall in the dark matter gravitational potential well. This process can be responsible for a significant
modification to the slope of the dark matter density profile at small radii, leading to large core radii.
In the model of Ref.~\cite{elzant}, the final dark matter density distribution can be approximated by a profile such as:

\begin{equation}\label{eq:burk}
g_{\rm Burk}(y)=\frac{1}{\left(1+x\right)\left(1+x^2\right)},
\end{equation}

\noindent which we refer to as the Burkert profile \cite{burkert}.  For comparison, in Fig.~\ref{fig:ANGULAR} in Sec.~\ref{sec:space}, where we compare the spatial distribution of gamma-ray emission for different models, we also show the radial profile of gamma-ray emission from a Burkert profile with our two reference substructure setups, but for simplicity we do not simulate this setup.  The radial distribution of gamma-ray flux for this profile versus an NFW profile differs only in the inner cluster regions ($R<0.5$ degrees for Coma).

Our choice of the particle dark matter models for the present study was motivated by considering a reasonable range of masses and two different dominant final state annihilation modes. In addition, the size of the pair annihilation cross section was fixed according to either theoretical or phenomenological arguments. In the interest of generality and in order to make our results easily reproducible and comparable to previous work, we do not pick specific theoretical particle physics frameworks, but rather we specify the dominant final state, the particle mass and its pair annihilation rate. This allows one to completely determine the gamma-ray emission.

We consider a model with a relatively large mass ($m_{\rm WIMP}=110$ GeV), a dominant $W^+W^-$ final state annihilation mode, and a cross section corresponding to what is expected, for that mass, for a wino-like neutralino (i.e. for the supersymmetric fermion corresponding to the SU(2) gauge boson), namely $\langle\sigma v\rangle=1.5\times 10^{-24}\ {\rm cm}^3/{\rm s}$. We choose the mentioned value for the WIMP mass for two reasons: (1) we want to use the gamma-ray spectrum resulting from a $W^+W^-$ final state annihilation mode, which forces us to consider $m_{\rm WIMP}>M_W\approx80.4$ GeV, and (2) we want a sizable gamma-ray flux, which forces us to consider a relatively light mass. The choice of $110$ GeV serves both the purpose of avoiding fine-tuning with the $W$ threshold and of being heavy enough to contrast it to our second WIMP setup choice, described below. We call this model H, for high mass, and we assume that such a WIMP has a number density in accord with the cold dark matter abundance thanks to either non-thermal production \cite{nth} or to a modified cosmological expansion at the WIMP freeze-out \cite{mod}. 

Our second WIMP setup is a low mass model (L), featuring $m_{\rm WIMP}=40$ GeV (the lightest mass compatible with grand unified gaugino masses), a dominant $b\bar b$ final state, and a pair annihilation cross section approximately in accord with what is expected for thermal production of cold dark matter,  $\langle\sigma v\rangle=6\times 10^{-26}\ {\rm cm}^3/{\rm s}$. While approximately $\Omega_\chi h^2\approx (3\times 10^{-26}\ {\rm cm}^3/{\rm s})/\langle\sigma v\rangle$ \cite{dmreviews}, the scatter to that relation for instance in supersymmetric models (e.g. from resonant annihilation channels) justifies a slightly larger value, which enhances our predicted gamma-ray fluxes.

Having specified the setup for both the dark matter density distribution and particle properties, we can compute the differential gamma-ray yield (number of photons per unit energy, time and surface) as the following integral over the line of sight:
\be
\frac{{\rm d}\Phi_\gamma}{{\rm d}E_\gamma}=\int_{\rm l.o.s.}{\rm d}l\ \frac{\langle\sigma v\rangle}{4\pi}{\mathcal N}_{\rm pairs}(r(l))\frac{{\rm d}N^f_\gamma}{{\rm d}E_\gamma} (E_\gamma),
\ee
where ${\rm d}N^f_\gamma/{\rm d}E_\gamma(E_\gamma)$ stands for the differential  gamma-ray yield per annihilation for final state $f$, as resulting from the Pythia \cite{pythia} Monte Carlo simulations implemented in the DarkSUSY code \cite{ds}. 

\begin{figure}[t]
\begin{center}
\includegraphics[width=16.5cm,clip]{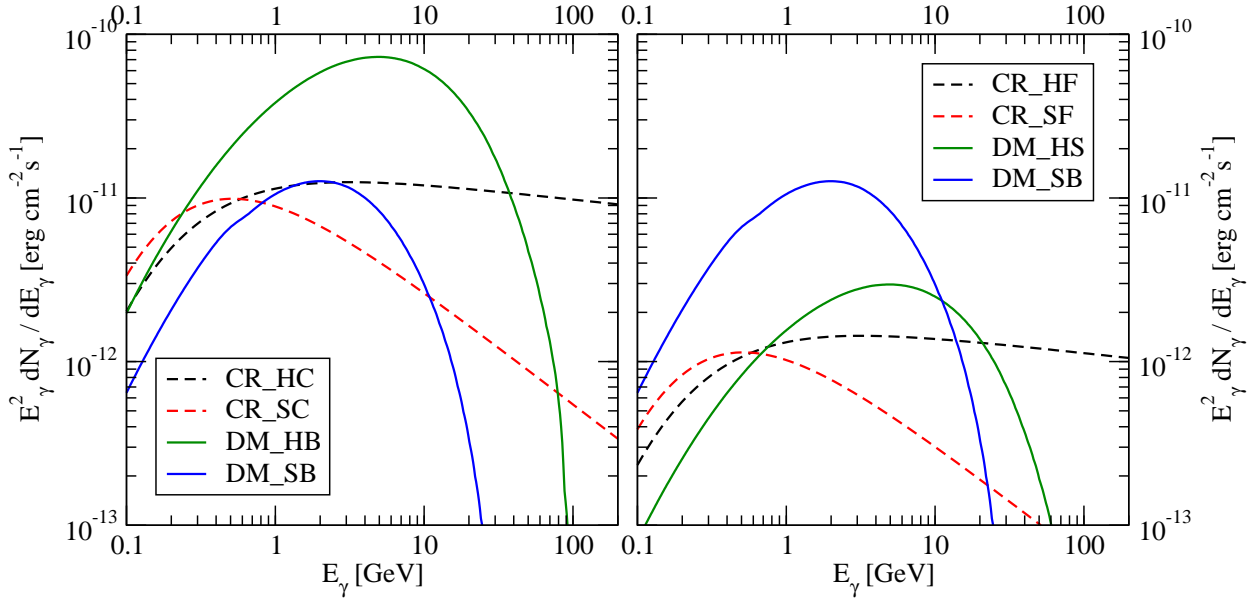}\\
\caption{Differential gamma-ray spectrum times gamma-ray energy squared (i.e. spectral energy density), for the various models considered in the present analysis. ``CR'' indicates cosmic ray models, while ``DM'' the dark matter annihilation emission. See the text and Tab.~\ref{tab:cr_models} and \ref{tab:grtab} for the meaning of the other labels and the associated input parameters. \label{fig:SPECTRA}}
\end{center}
\end{figure}

We summarize the dark matter model parameters in Tab.~\ref{tab:grtab}. The resulting gamma-ray fluxes are summarized instead in the second column of Tab.~\ref{tab:spec}. The various gamma-ray spectra of our cosmic ray and dark matter benchmark models are shown in Fig.~\ref{fig:SPECTRA}.

\begin{table}
\begin{tabular}{cccccc}
\hline
Model & Mass & Final & $\langle\sigma v\rangle$ & $f_s$ & $\Delta^2$ \\
ID & $m_{\rm WIMP}$/GeV & State & $[{\rm cm}^3/{\rm s}]$ & &  \\
\hline
DM\_HB & 110 & $W^+W^-$ & $1.5\times 10^{-24}$ & 0.5 & $7\times 10^6$ \\
DM\_HS & &  &  & 0.2 & $7\times 10^5$ \\
\hline
DM\_LB & 40 & $b\bar b$ & $6\times 10^{-26}$ & 0.5 & $7\times 10^6$ \\
DM\_LS & &  &  & 0.2 & $7\times 10^5$ \\
\hline
\end{tabular}
\caption{Input parameters for the dark matter models considered here. The quantity $f_s$ indicates the ratio of the total mass in subhalos over the total virial mass, while $\Delta^2$ stands for the weighed enhancement in the number density of dark matter particle pairs due to subhalos. \label{tab:grtab}}
\end{table}

\subsection{Fermi Simulation Setup}\label{sec:glastsim}
We produced simulated Fermi observations using the Fermi-LAT observation simulator tool, {\tt gtobssim}, in the Fermi Science Tools package (v9r7) \cite{sciencetools}.  We run simulations for the specific case of the Coma cluster in terms of the cluster distance, mass, and size.  As discussed above, we choose two spectral and two spatial models for the gamma-ray emission from both cosmic rays and dark matter annihilation which encompass the ranges expected for these sources.  As shown below (see Sec.~\ref{sec:specs}), these models cover a wide range of gamma-ray fluxes from clusters and, therefore, represent a reasonable range in the possible signal-to-noise for clusters detectable by Fermi.  The range and variation in expected gamma-ray flux from known clusters is considered in detail in Sec.~\ref{sec:cat}.  Our simulated models simply provide benchmarks of what could be seen for observed clusters with similar statistics.

The simulations were run in the default scanning mode with the Pass 6 instrumental source response functions (P6\_V1\_SOURCE).  For each cluster simulation, data files defining the cluster spectrum and images defining the spatial distribution were fed to {\tt gtobssim} (see Sec.~\ref{sec:cr} and \ref{sec:dm} for model definitions).  To include the extragalactic diffuse background emission, we simulate an isotropic source with the power-law spectral parametrization \cite{burkert} found in the analysis of the EGRET data \cite{egreteg}.  We note that the Fermi background may be lower if a significant fraction of the extragalactic background is resolved as AGN \cite{agn1,agn2,agn3,agn4,agn5}. 

\begin{figure}[!h]
\begin{center}
\includegraphics[width=16cm,clip]{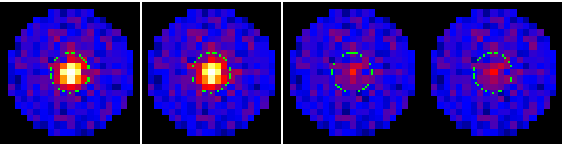}\\
{\bf CR\_HC\hspace*{2.3cm} CR\_SC\hspace*{2.3cm} CR\_HF\hspace*{2.3cm} CR\_SF}\\
\caption{ Shown from left to right are images of simulations of five year Fermi observations of the Coma cluster for models CR\_HC, CR\_SC, CR\_HF, and CR\_SF, including the extragalactic diffuse background. Images are 20 degrees across and binned to have 1 degree pixels. The color bar is the same in all cases. \label{fig:crimages}}
\end{center}
\end{figure}

\begin{figure}[!h]
\begin{center}
\includegraphics[width=16cm,clip]{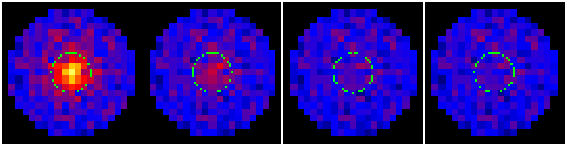}\\
{\bf DM\_HB\hspace*{2.3cm} DM\_LB\hspace*{2.3cm} DM\_HS\hspace*{2.3cm} DM\_LS}\\\caption{ Shown from left to right are images of simulations of five year Fermi observations of the Coma cluster for models DM\_HB, DM\_LB, DM\_HS, and DM\_LS, including the extragalactic diffuse background. Images are 20 degrees across and binned to have 1 degree pixels.  The color bar matches the color bar in Fig.~\ref{fig:crimages}. \label{fig:dmimages}}
\end{center}
\end{figure}
We simulate a one year observation for each combination of our benchmark spectral and spatial models for gamma-ray emission from cosmic rays and dark matter annihilation.  We abbreviate these eight combinations as follows: DM\_HS for a high dark matter mass and a smooth dark matter distribution, DM\_LS for a low dark matter mass and smooth distribution, DM\_HB for a high mass and a dark matter distribution with significant substructure/boost factor, DM\_LB for a low mass and a significant substructure boost, CR\_HF for a hard cosmic ray spectrum and a flat cosmic ray spatial distribution, CR\_SF for a soft spectrum and flat distribution, CR\_HC for a hard spectrum and a cuspy cosmic ray distribution, and finally, CR\_SC for a soft spectrum and cuspy distribution.  Each simulated source is normalized to have the total model predicted flux for the given spectral and spatial models integrated over a region of 10.5 degree radius (the extent of our input images to {\tt gtobssim}).  The input fluxes and the number of simulated source photons are given in Tab.~\ref{tab:spec}. Additionally we simulate five year long observations of all models to test the improvement to our fits with higher statistics, as discussed below.  In Fig.~\ref{fig:crimages} and ~\ref{fig:dmimages} we show images of the five year cosmic ray and dark matter simulations, respectively, binned to one degree and including the extragalactic background.  The range in fluxes of these models is apparent.  In particular, with little boost from substructure (DM\_HS and DM\_LS) the dark matter models are very faint.  In addition, the spatial distribution of the gamma-ray emission from dark matter (DM\_HB) appears more extended than from cosmic rays (see also Sec.~\ref{sec:space}). We remark that we neglect here additional non-galaxy-cluster point sources expected, on average, in a 20 degree field with fluxes comparable or larger than what we predict for our cluster emission models. Specifically, estimates of the number of high-latitude gamma-ray point sources based upon the extrapolation of EGRET results and/or blazar models \cite{Ciprini:2003nwa}, or on the actual Fermi-LAT early results \cite{fermiblazars} indicate that one would expect within a 10 deg radius region (what we consider in our figures) $\sim1.5$ sources with a flux at or above the boosted high-mass DM setup DM\_HB ($\gtrsim50\times 10^{-9}\ {\rm cm}^{-2}{\rm s}^{-1}$).

\section{Deducing the Origin of Gamma Rays from Galaxy Clusters}\label{sec:origin}

\subsection{The Optimal Angular Region}\label{sec:angle}

Prior to embarking on the spectral and spatial analysis of our gamma-ray simulations, and of addressing the question of the potential of Fermi to discriminate between a gamma-ray emission in galaxy clusters originating from dark matter versus cosmic rays, in the present Section we investigate the optimal angular cuts from a theoretical standpoint. We show results for our one-year simulations for all of our cosmic ray and dark matter models, as well as for the extragalactic diffuse cosmic-ray background. 

\begin{figure}[t]
\begin{center}
\includegraphics[width=16.5cm,clip]{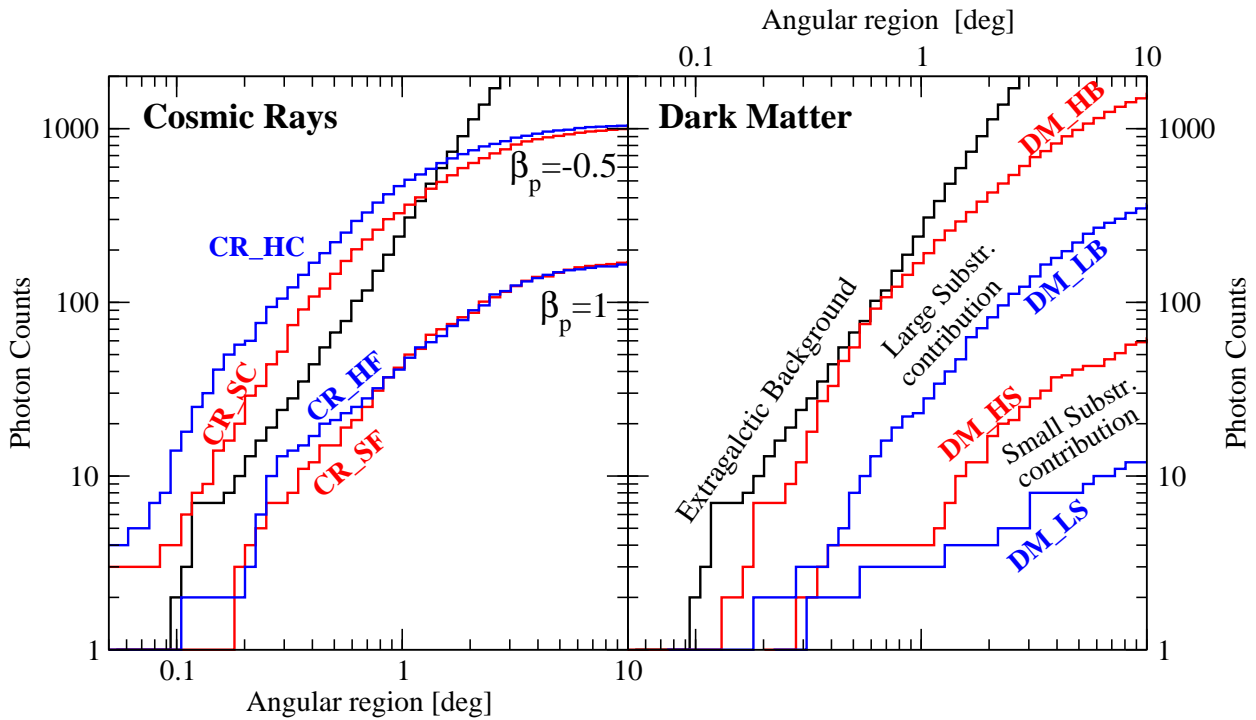}\\
\caption{The results of our simulations of one-year observations, binned by angular regions. We take as a reference case that of the Coma cluster, and we are summing over all photon counts in our simulations with energies above 100 MeV. The black line indicates the estimated extragalactic background. \label{fig:counts}}
\end{center}
\end{figure}
Given the low statistics of photon counts, we decided to include all photons with a reconstructed energy above 0.1 GeV. Fig.~\ref{fig:counts} shows the number of photon counts inside given angular regions, specified on the x-axis, for the various models. The left panel shows the four cosmic ray cases, while the right panel the dark matter annihilation induced gamma-ray signal. In both panels, for reference, we also show our simulated extragalactic gamma-ray background. A naive by-eye signal-to-background inspection would indicate a small angular region (a few tenths of a degree) as the optimal choice. However, given the fact that we will actually be able to subtract with some efficiency the extragalactic background, it makes sense to investigate the ratio of the signal to the noise, or square root of the background. In addition, the sheer small number of photon counts if we chose a small angular region contain very little information.

\begin{figure}[t]
\begin{center}
\includegraphics[width=14.cm,clip]{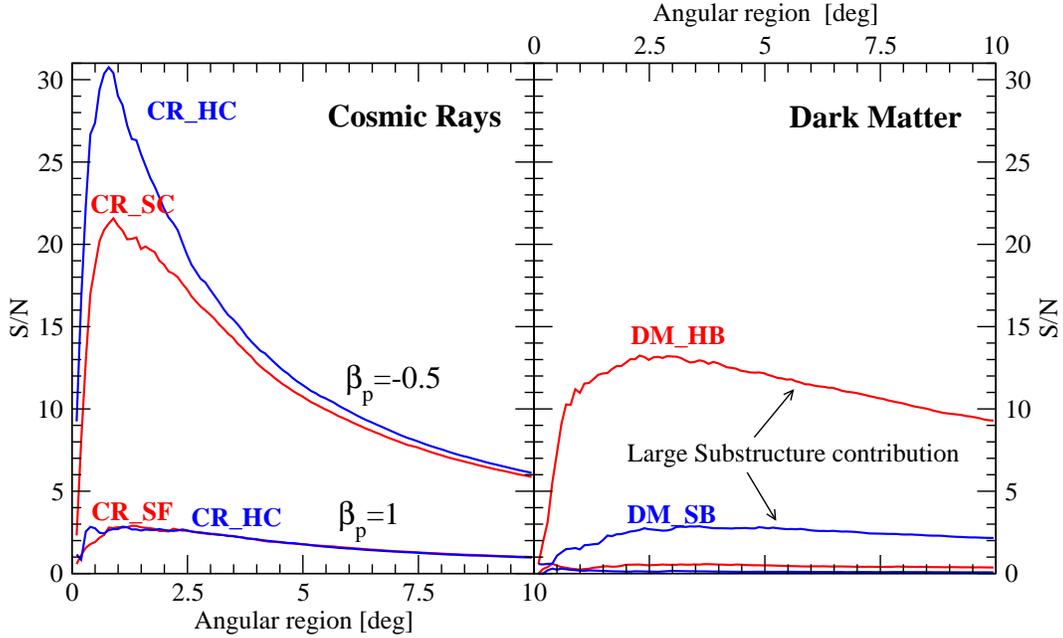}\\
\caption{The signal over square root of the background (S/N$\sim n\sigma$) in simulated one-year observations, as a function of the angular region, for all models under consideration here. \label{fig:SN}}
\end{center}
\end{figure}
Figure \ref{fig:SN} shows the ratio of the signal to the square root of the background, again for all models under consideration, and as a function of the angular region of interest. In the most luminous cosmic ray cases ($\beta_p=-0.5$), it appears that the signal-to-noise is maximized for a region of interest of around one degree. For such a choice, the left panel of Fig.~\ref{fig:counts} tells us that we would get around 500 photons per year above 0.1 GeV inside a one degree region, which is approximately 50\% of the overall photon flux. A choice of one degree seems thus optimal in the case of cosmic rays.

The brightest dark matter cases, on the other hand, feature a large and ``luminous'' substructure content, in particular at large radii, making the signal significantly more diffuse than in the cosmic ray case. Given the importance of differentiating between a point-like and a diffuse emission from clusters, it will be important to have a large enough region of interest. Additionally, the signal-to-noise for the dark matter cases is maximized in regions between 1.5 and 3.5 degrees. Again looking at Fig.~\ref{fig:counts}, we confirmed that the choice of a region of interest of 3 degrees appears to be optimal for a dark matter type signal.

In conclusion, we find that the analysis of the signal-to-noise and of the total photon counts would lead us to employ a region of interest of 3 degrees for the dark matter and of one degree for cosmic rays. In the interest of being sensitive to soft spectra, (the instrument Point Spread Function (PSF) would lead to the loss of most of the low energy photons out of an angular region of one degree) we decided to proceed with a 3 degree region for both cases.  For reference, we mark a 3 degree radius region in figs.~\ref{fig:crimages} and ~\ref{fig:dmimages}.

\subsection{Spectral Analysis}\label{sec:specs}

We examine the simulated spectra for our eight benchmark cluster models to investigate our ability to discriminate the gamma-ray spectra from cosmic rays versus dark matter annihilation.  We extract source spectra and response files, including the cluster plus extragalactic diffuse background, within a 3 degree radius of the cluster center using the tools {\tt gtbin} and {\tt gtrspgen}.  To account for the extragalactic diffuse emission, we extract a background spectrum from an annular region with an inner radius of 10 degrees and an outer radius of 12 degrees.  As is clear from Fig.~\ref{fig:counts}, this outer region contains very little cluster emission.  Spectra were fit using the XSPEC spectral fitting package \cite{xspec}.  Within XSPEC, the background spectrum is subtracted from the source spectrum after adjusting the BACKSCAL header keyword in the background file to account for the difference in area between the background spectrum and the source spectrum.

\begin{table}
\begin{tabular}{ccccccc}
\hline
Model & Flux ($>0.1$ GeV) & Source & $\alpha_p$ & $\chi_{\nu}$ & $m_{{\rm WIMP}}$ & $\chi_{\nu}$ \\
 & ($10^{-9}$ cm$^{-2}$s$^{-1}$) & Counts & & &(GeV) & \\
\hline
DM\_HB &54.7 &1823 &$2.27^{+0.08}_{-0.06}$ &\textbf{1.51} &$92.0^{+8.0}_{-16.1}$ &1.02 \\
DM\_LB &14.6 &431 &$2.87^{+1.05}_{-0.49}$ &1.01 &$<113.5$ &1.01 \\
DM\_HS &2.25 &73 &- &- &- &- \\
DM\_LS &0.597 &15 &- &- &- &- \\
\hline
CR\_HC &37.7 &1161 &$2.39^{+0.08}_{-0.07}$ &1.12 &$24.8^{+3.1}_{-4.6}$ &1.11 \\
CR\_SC &42.9 &1188 &$2.89^{+0.14}_{-0.13}$ &0.95 &$<11.7$ &\textit{0.97} \\
CR\_HF &6.76 &187 &$3.26^{+1.52}_{-0.73}$ &1.01 &$<54.3$ &\textit{1.01} \\
CR\_SF &7.71 &208 &$4.05^{+2.10}_{-1.11}$ &1.01 &$<40.0$ &\textit{1.02} \\
\hline
\end{tabular}
\caption{ Summary of one year simulations and their spectral fits, for the specific case of the Coma cluster. The second column lists the total flux input in to the simulations ($>100$ MeV), and the third column gives the total number of simulated cluster photons (all angles).  Columns 4 and 5  give the best-fit slope and reduced $\chi^2$ for the CLUSTERCR fits, while columns 6 and 7 list the best-fit particle mass and reduced $\chi^2$ for the DMFIT fits.  Upper/lower limits on spectral parameters refer to 90\% confidence limits; all other errors are 1 sigma.  Reduced chi-squared ($\chi_\nu$) in bold indicate a fit probability of less than 1\% and in italics indicate that the best fit is found for the DMFIT lower mass limit of 10 GeV. \label{tab:spec}}
\end{table}

XSPEC allows one to include custom user models, and we use this feature to include models for the gamma ray spectrum from both cosmic rays and dark matter annihilation.  For the dark matter spectrum, we use the routine DMFIT, which we presented in ref.~\cite{ourgc}.  DMFIT is a tool that provides the gamma-ray flux from generic WIMP pair annihilation (i.e. from dark matter particles with specified mass and branching ratios into Standard Model final state annihilation modes). DMFIT is based on the same set of Monte Carlo simulations used in DarkSUSY \cite{ds} and incorporates a wide variety of annihilation modes. Two data files contain the Monte Carlo simulation results giving the differential and integrated gamma-ray fluxes at given energies. The simulation results are then interpolated given the dark matter particle mass and annihilation final states supplied by the fitting routine. We additionally include the $e^+e^-$ channel, where gamma-rays are radiated in the final state via internal bremsstrahlung. The $e^+e^-$ channel -- presently not included in the DarkSUSY code -- is relevant for various non-supersymmetric WIMP models, including the Kaluza-Klein dark matter of Universal Extra Dimensions \cite{kkdm} (see Ref.~ \cite{nicole} for model-independent limits on the annihilation cross section of dark matter to $e^+e^-$). 

DMFIT consists of two data files and one Fortran routine, and the code is publicly available from the authors upon request. DMFIT essentially reverse-engineers the use of the DarkSUSY package for the computation of gamma-ray spectra: given an observed gamma-ray spectrum, DMFIT allows one to fit for the best matching particle dark matter mass, its pair-annihilation rate and its branching ratios. In conjunction with virtually any fitting package, like XSPEC and the {\tt gtlike} routine in the Fermi Science Tools, DMFIT can be used to reconstruct confidence level ranges for the mentioned particle dark matter properties.

While the {\tt Pythia} Monte Carlo simulations extend down to a WIMP mass of 10 GeV, DMFIT allows one to extrapolate to lower masses. Very light WIMPs have been recently shown to be relevant even in the context of supersymmetry \cite{Profumo:2008yg}, and they can possibly play a role in explaining the puzzling DAMA/LIBRA signal \cite{dama}.  We note that in the current XSPEC version of DMFIT, the default lower limit on the dark matter particle mass is 10 GeV.  For the spectral fits shown below with only upper limits on the particle mass, the best fit typically saturates at the 10 GeV limit.  

The spectrum from cosmic ray interactions is modeled with the simple analytic form given in Eq.~(\ref{eq:crspec}), and included in XSPEC using a Fortran routine as a model named CLUSTERCR.  Both XSPEC models are available from the authors upon request, and both models are currently being incorporated into the Fermi-LAT Science Tools as part of the {\tt gtlike} likelihood fitting tool.

Tab.~\ref{tab:spec} summarizes the spectral fits to simulations of one year observations of the Coma cluster.  For each simulated cluster spectrum, we fit both a cosmic ray and a dark matter model, regardless of the input model, to compare how well we can distinguish these two scenarios based on the spectrum.  For simplicity, in the case of the cosmic ray simulations, we assume a $b\bar b$ final state for the DMFIT spectral model, while for the dark matter simulations, we use the dominant final state ($b\bar b$ for DM\_L and $W^+W^-$ for DM\_H).  The overall spectral shape of the $b\bar b$ final state is quite similar to most other final states \cite{Cesarini:2003nr,Profumo:2005xd,ds}, including $W^+W^-$, but we refer the reader to ref.~\cite{ourgc} for a discussion of the systematic affects of the assumption of final state on the reconstruction of dark matter particle properties.  Columns 4 and 5 of Tab.~\ref{tab:spec} give the best-fit slope and reduced $\chi^2$ for the CLUSTERCR fits, while columns 6 and 7 list the best-fit particle mass and reduced $\chi^2$ for the DMFIT fits.

\begin{figure}[t]
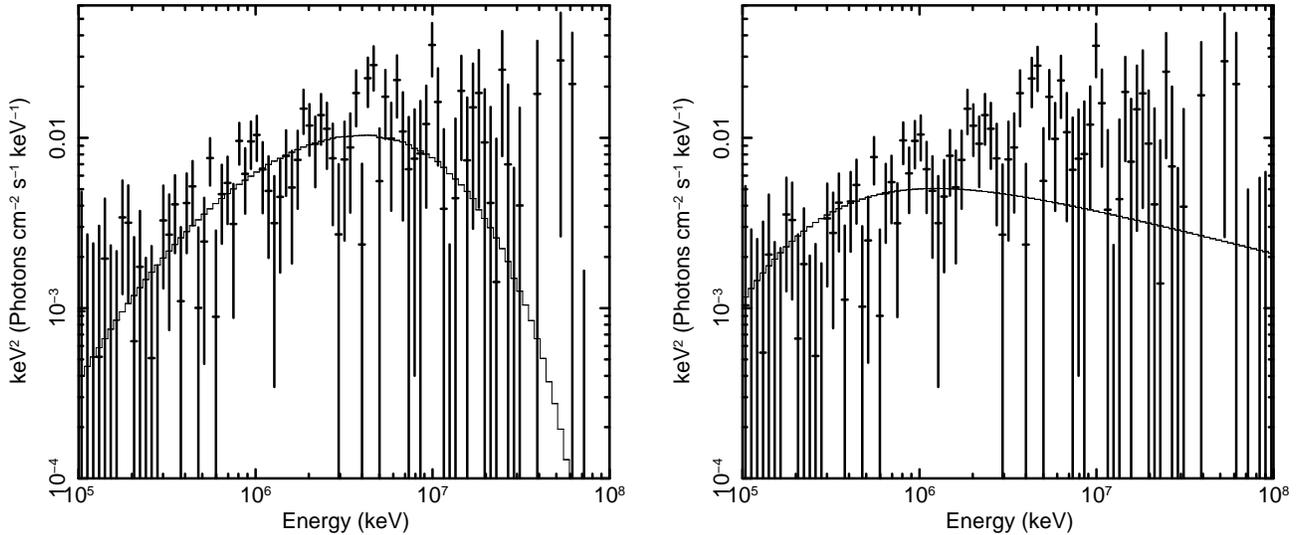

\begin{center}
\mbox{\hspace*{-0.3cm}\includegraphics[height=8.5cm,clip,angle=-90]{dmopts_dmfit_spec.ps}\quad \includegraphics[height=8.5cm,clip,angle=-90]{dmopts_cr_spec.ps}}\\
\caption{Shown are fits to the one year simulated spectrum for model DM\_HB ($m_{{\rm WIMP}} = 110$ GeV), for the specific case of the Coma cluster.  The left panel shows the best-fit dark matter model using DMFIT with a $W^+W^-$ final state.  The right panel shows to same simulated data and best-fit cosmic ray spectrum using CLUSTERCR.  The cosmic ray model is ruled out at better than 99.9\% confidence. \label{fig:dmoptsspec}}
\end{center}
\end{figure}

For a higher dark matter mass and a reasonably bright source, as is the case for DM\_HB ($m_{{\rm WIMP}} = 110$ GeV), we find a good fit to a dark matter spectrum, but we cannot get a good fit to a cosmic ray spectrum.  Fig.~\ref{fig:dmoptsspec} shows a comparison of the best-fit DMFIT (left panel) and CLUSTERCR (right panel) models for this simulation compared to the data.  In this case, the spectrum is only consistent with a dark matter interpretation; the best-fit dark matter particle mass is low by $\sim 15$\% but consistent with the true mass within a couple of sigma.  On the other hand, our low dark matter mass model DM\_LB ($m_{{\rm WIMP}} = 40$ GeV), which is also significantly fainter, can be well fit by either a cosmic ray spectrum with a fairly steep slope or with a dark matter spectrum.  

A similar result is seen when considering the the simulated cosmic ray models.  In general, these are consistent with either a cosmic ray spectrum or a dark matter spectrum with a low mass dark matter particle.  The cuspy cosmic ray distribution gives fluxes 5-10 times higher than a flatter distribution, and in these cases the possible dark matter mass is limited to be quite low ($24.8^{+3.1}_{-4.6}$ GeV for a hard cosmic ray spectrum and a 90\% upper limit of 11.7 GeV for a soft cosmic ray spectrum).  Even for the fainter CR\_HF and CR\_SF models, we find a 90\% upper limit on the possible dark matter mass of $40-55$ GeV.  Interestingly, the cosmic ray spectral slope, $\alpha_p$, is overestimated in all cases, though the best fits are within a couple sigma of the true slope.  This is most likely due to the hard spectrum of the extragalactic background emission.  As the dark matter particle mass is also somewhat underestimated for DM\_HB, it appears that the background subtraction leads to a general softening of the source spectrum.  The effect of the background can be mitigated and the cosmic ray spectral slope recovered to good accuracy if a smaller energy rage extending only up to 10 GeV is used in the spectral fits.  However, with such a low high energy cutoff, a high mass dark matter particle model, like DM\_HB, is indistinguishable from a hard cosmic ray spectrum.  We, therefore, recommend using a large energy range ($\sim 0.1-150$ GeV) to investigate the dominant source of the emission but a lower high energy cutoff ($\sim0.1-10$ GeV) to determine the model parameters.

\begin{table}
\begin{tabular}{ccccccc}
\hline
Model & Flux ($>0.1$ GeV) & Source & $\alpha_p$ & $\chi_{\nu}$ & $m_{{\rm WIMP}}$ & $\chi_{\nu}$ \\
 & ($10^{-9}$ cm$^{-2}$s$^{-1}$) & Counts & & &(GeV) & \\
\hline
DM\_HB &54.7 &9171 &$2.10^{+0.02}_{-0.02}$ &\textbf{2.94} &$100.5^{+5.5}_{-5.7}$ &0.96 \\
DM\_LB &14.6 &2343 &$2.47^{+0.13}_{-0.15}$ &1.03 &$31.7^{+8.0}_{-4.0}$ &0.97 \\
DM\_HS &2.25 &368 &$>2.19$ &0.98 &$<200$ &1.00 \\
DM\_LS &0.597 &100 &- &- &- &- \\
\hline
CR\_HC &37.7 &5819 &$2.23^{+0.03}_{-0.03}$ &1.11 &$29.2^{+2.0}_{-1.0}$ &\textbf{2.02} \\
CR\_SC &42.9 &5990 &$2.76^{+0.06}_{-0.06}$ &0.99 &$<11.1$ &\textit{\textbf{1.40} }\\
CR\_HF &6.76 &977 &$2.66^{+0.29}_{-0.21}$ &0.99 &$24.1^{+8.0}_{-14.1}$ &1.03 \\
CR\_SF &7.71 &1145 &$3.32^{+0.25}_{-0.61}$ &0.92 &$<13.0$ &\textit{0.98} \\
\hline
\end{tabular}
\caption{ Summary of five year simulations and their spectral fits. The second column lists the total flux input in to the simulations ($>100$ MeV), and the third column gives the total number of simulated cluster photons (all angles).  Columns 4 and 5 give the best-fit slope and reduced $\chi^2$ for the CLUSTERCR fits, while columns 6 and 7 list the best-fit particle mass and reduced $\chi^2$ for the DMFIT fits.  Upper/lower limits on spectral parameters refer to 90\% confidence limits; all other errors are 1 sigma.  Reduced chi-squares in bold indicate a fit probability of less than 1\% and in italics indicate that the best fit is found for the DMFIT lower mass limit of 10 GeV. \label{tab:spec5}}
\end{table}

\begin{figure}[t]
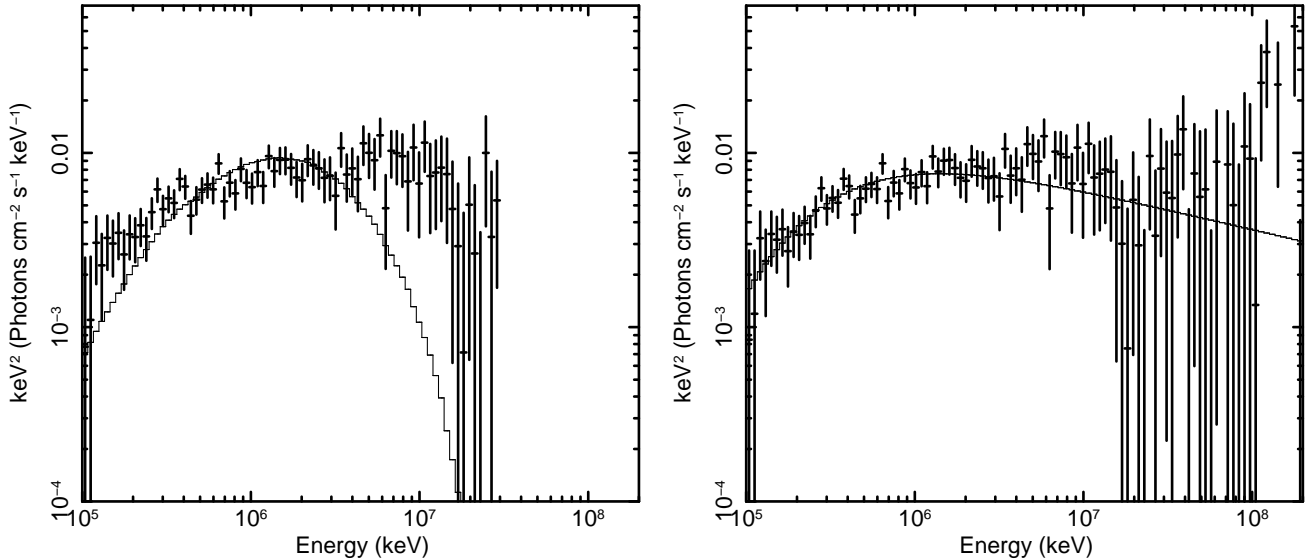

\begin{center}
\mbox{\hspace*{-0.3cm}\includegraphics[height=8.5cm,clip,angle=-90]{crpesp5yr_dmfit_spec.ps}\quad \includegraphics[height=8.5cm,clip,angle=-90]{crpesp5yr_cr_spec.ps}}\\
\caption{ Fits to the five year simulated spectrum for model CR\_HC ($\alpha_p = 2.1$).  The left panel shows the best-fit dark matter model using DMFIT with a $b\bar b$ final state, and the right panel shows the best-fit cosmic ray spectrum using CLUSTERCR.  The dark matter model is ruled out at better than 99.9\% confidence.  Note that in the left panel data points above the best-fit particle mass are not shown as the flux here is predicted to be zero. \label{fig:crpespspec}}
\end{center}
\end{figure}

As Fermi is expected to have at least a five year mission lifetime, we also consider the improvement in the spectral constraints for five year simulations of the same models; the results are shown in Tab.~\ref{tab:spec5}.  With deeper observations, even a low mass dark matter particle model is ruled out for the brighter cosmic ray models, CR\_HC and CR\_SC, and for the fainter models, CR\_HF and CR\_SF, the upper limit on the particle mass is significantly decreased requiring a very light dark matter particle.  As an example, we show in Fig.~\ref{fig:crpespspec} the best-fit DMFIT (left panel) and CLUSTERCR (right panel) models to the five year simulated spectrum of CR\_HC ($\alpha_p = 2.1$).  A dark matter model is clearly inconsistent with the spectral data.  As noted above, here we have only considered a $b\bar b$ final state, a reasonable assumption as it has a similar spectral shape to most other potential final states.   However, we note that if we fit a $\tau^+\tau^-$ final state, which has a significantly harder gamma-ray spectrum than  $b\bar b$, we get a much worse fit to the cosmic ray simulations.  With five years of data, the best-fit cosmic ray slopes for the CLUSTERCR model are closer to the input values, but they are still too high by $\sim1 \sigma$ for the soft spectral model and $\sim3 \sigma$ for the hard spectral model.  Again, this offset can be removed by using a smaller energy range (0.1-10 GeV) in the spectral fit to reduce the effects of the background.

For the dark matter simulations DM\_HB and DM\_LB, the longer exposure time significantly reduces the errors on the particle mass estimates, decreasing both the systematic shift to lower masses and the statistical errors.  The simulated high particle mass spectrum is again very inconsistent with a cosmic ray-type spectrum.  However, the low mass model with its moderate flux is still well fit by a cosmic ray model.

Unfortunately, the simulations of emission from dark matter annihilation with a smooth dark matter spatial distribution (i.e.~little substructure) have very low fluxes and low S/N even with five years of data.  For these simulations (DM\_HS and DM\_LS), we cannot get good constraints on the spectral model.  The best case with the smooth spatial distribution is the five year simulation of DM\_HS, and as can be seen in Tab.~\ref{tab:spec5}, we can only derive a lower limit on the possible cosmic ray slope and an upper limit on the possible dark matter particle mass.

In summary, assuming a dominant cosmic-ray or dark-matter origin, gamma ray emission from dark matter annihilation with a high dark matter particle mass ($m_{{\rm WIMP}}>50$ GeV) can be distinguished from a cosmic ray spectrum even for fairly faint sources (one year data for CR\_HF and CR\_SF, for example).  Distinguishing a cosmic ray spectrum from a low dark matter particle mass is much more difficult and requires deep data and/or a bright source.

While our models show that either cosmic rays or dark matter annihilation 
can dominate the gamma-ray emission from clusters, a mix of the two is 
another likely scenario.  As a final test, we simulate two clusters with 
emission from both cosmic rays and dark matter with equal fluxes for five 
years of observing time.  The first simulation combines emission from 
models DM\_HB and CR\_SC each normalized to have a flux of $2.5 \times 
10^{-4}$ photons m$^{-2}$ s$^{-1}$; the second simulation combines 
emission from DM\_LB and CR\_HC each with a flux of $1.3 \times 10^{-4}$ 
photons m$^{-2}$ s$^{-1}$.  Note that the total cluster fluxes from both 
sources of emission are then within the range of the original models.  The 
first case is more optimistic combining a high mass dark matter particle 
with a soft cosmic ray spectrum, while the second case has about half the 
total flux and combines a low particle mass with a hard cosmic ray 
spectrum.  We do not simulate fainter mixed models, as we have already 
demonstrated the difficulty of distinguishing the type of emission for 
faint sources.

For the more optimistic case of a high mass dark matter particle and a 
soft cosmic ray spectrum, we find that neither a cosmic ray only spectral model 
($\chi_{\nu} = 1.35$ for 94 DOF) nor a dark matter only model ($\chi_{\nu} = 2.03$ for 94 DOF) provide a good fit.  Fitting to a combined dark matter 
and cosmic ray spectral model, however, gives a good fit ($\chi_{\nu} = 0.87$ 
for 92 DOF) with the normalizations of both components non-zero at $> 3 \sigma$.  Here the DM particle mass is underestimated and the cosmic ray slope overestimated by $2-3 \sigma$ with $m_{{\rm WIMP}} = 87^{+8}_{-4}$ GeV and $\alpha_p = 3.5\pm0.4$.  If the dark matter particle mass is known from other astronomical observations or direct detection experiments, then the errors can be significantly reduced.  Fixing the dark matter mass at the true value, we find $\alpha_p = 3.1^{+0.1}_{-0.2}$ and the normalizations of the cosmic ray and dark matter components are both determined to $\sim15$\%.  The results for the second mixed simulation with a low particle mass and hard cosmic 
ray spectrum unfortunately show that one cannot distinguish the source of the emission.  Comic ray only, dark matter only, and mixed models all give good fits to the data even with the dark matter particle mass fixed at its true value.

Overall, we find, for bright enough sources, that if the dark matter 
contribution to the cluster flux is significant and the particle mass is 
not very low ($m_{WIMP} > 40$ GeV) the presence of a dark matter component 
can be seen even in the presence of a significant gamma-ray flux from 
cosmic rays.  The reconstruction of the model parameters is more difficult in this case, but can be significantly improved if something about the dark matter mass in known.

\subsection{Hardness Ratios}

In addition to the spectral analysis, we also investigated the use of simple energy band ratios, hardness ratios, to differentiate the gamma-ray signal of cosmic rays and dark matter.  Hardness ratios give a quick estimate of the slope/shape of the spectrum and can be used as a rough spectral analysis for lower photon count data (for instance, dimmer sources or shorter observing time) when it is not possible to accurately estimate the flux in more than a couple of energy bins.

We defined two energy bands hard (H) and soft (S), and our hardness ratio is defined as HR = (H -- S)/(H + S).  A couple of considerations were used in defining these bands.  We chose a lower energy limit of $\sim 150$ MeV so that the expected cosmic ray spectrum is a simple power law (e.g.~away from threshold effects).  We also require that the hard band have at most a factor of $10$ difference in photon counts from the soft band for the range of models under consideration.  The energy bands we chose are S $=0.15 \leq E < 0.7$ GeV and H $=E \geq 0.7$ GeV.  This particular choice of bands has the additional convenient feature that the expected ratio, HR, is negative for cosmic rays and positive for dark matter.

As for the spectral analysis of our Fermi simulations, we used a $3$ degree radius source region and an annulus from $10$ to $12$ degrees to estimate the background.  The photon counts in the background region are subtracted from the counts in the same energy band in the $3$ degree source region, after accounting for the difference in area.  We compared this annular background signal to a simulation of the background in the source region, and they match very well.

\begin{figure}[t]
\begin{center}
\includegraphics[width=16.5cm,clip]{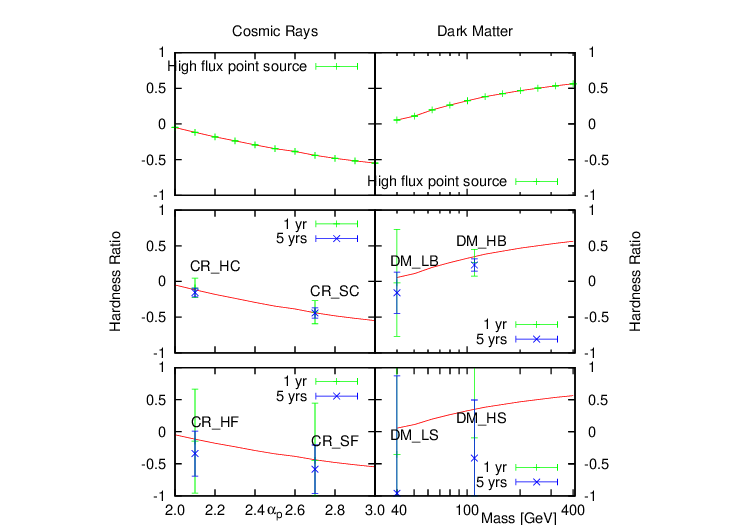}\\
\caption{The hardness ratio for cosmic ray and dark matter models.  Indicating with H the total number of (hard) photons with energies above 0.7 GeV, and with S those (soft) in the energy band between 0.15 and 0.7 GeV, we define here as {\em hardness ratio} the quantity (H-S)/(H+S). The top plots are for high flux point sources without background, used as reference results.  The middle and bottom rows of plots show the results for the simulated Fermi cluster observations of Coma, including the uncertainties due to the background subtraction.  The red line in each plot is the same as in the top plots (just connected, not a best fit). \label{fig:hr_big}}
\end{center}
\end{figure}


To find the expected hardness ratios for the different spectra after including the Fermi instrument response, we ran simulations of point sources for cosmic rays ($\alpha_p$ between $2$ and $3$) and dark matter (neutralino mass between $40$ and $400$ GeV, b-$\bar{\textrm{b}}$ final state), without any background and set to very high fluxes (top panel of Fig.~\ref{fig:hr_big}).  For all of these simulations, we took $10^5$ photons.  Here we indeed see that the expected ratio, HR, is negative for cosmic rays and positive for dark matter.  Also, at least with such good statistics, the ratio correlates well with $\alpha_p$ or the dark matter mass.  The results of our cluster simulations agree well with these results.  The hardness ratios for our eight simmulated cluster models, after background subtraction as described above, are plotted in the middle and lower panels of Fig.~\ref{fig:hr_big}, for 1 and 5 years of data.

All of these simulations agree well within errors with the results from the simple high-flux analysis, though the errors are large for the fainter cases.  
While it can be difficult to tell, for example, a lower dark matter mass from a harder cosmic ray spectrum, these results show that it is possible to use this simple two energy band ratio for distinguishing some cosmic ray and dark matter models.  
The hardness ratio also correlates well with spectral index $\alpha_p$ or the dark matter mass.  So, for instance, if it is known that a gamma-ray source is mostly dark matter in origin, this ratio provides an indication of the particle mass.  From comparing the simulations both with and without background we find that significant errors come from the uncertainty in the background and its subtraction.  Accurate modeling of the diffuse gamma-ray background could significantly improve this analysis.

\subsection{Spatial Extension of the Gamma-ray Emission}\label{sec:space}
\begin{figure}[t]
\begin{center}
\includegraphics[width=16.5cm,clip]{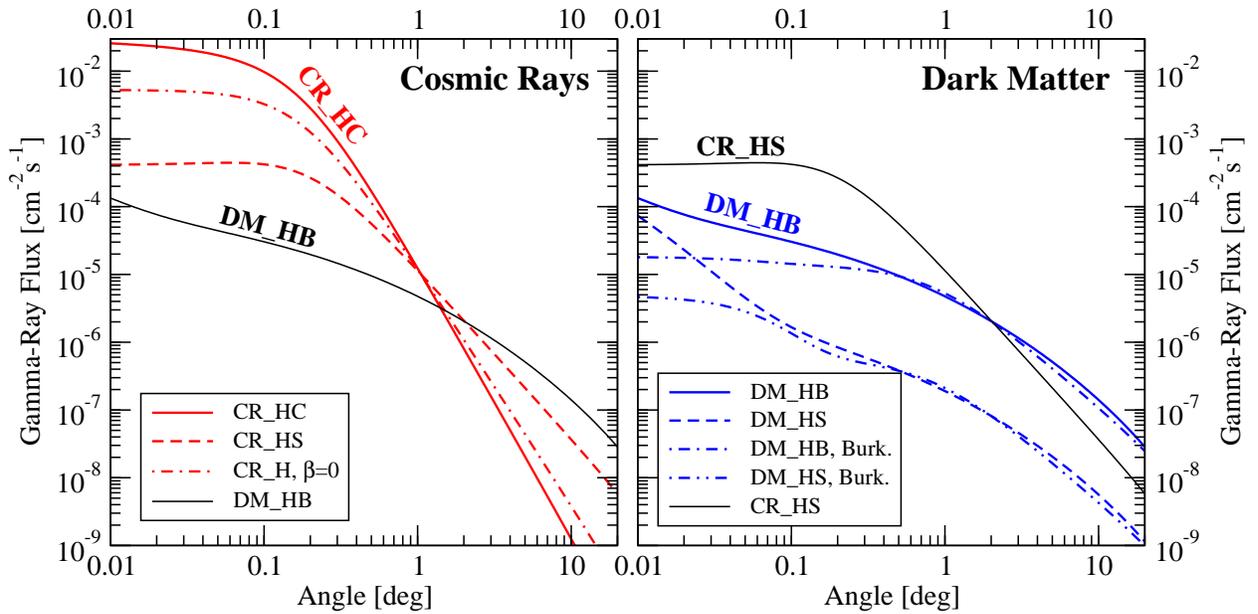}\\
\caption{The gamma-ray emission as a function of the angular distance from the center of the Coma cluster, for cosmic rays (left) and for dark matter (right). The first two lines in each panel correspond to our reference setups, and in both panels the thin solid black lines offer a comparison for dark matter (left) and cosmic rays (right). \label{fig:ANGULAR}}
\end{center}
\end{figure}

In addition to their spectra, the gamma-ray emissions from cosmic rays and dark matter annihilation in clusters are expected to have different spatial distributions.  With sufficient signal, cluster emission from either source may also be detectable as extended, which could distinguish a cluster signal from the emission from e.g. a nearby AGN.  In Fig.~\ref{fig:ANGULAR}, we show the gamma-ray flux versus angular distance from the cluster center for a range of models including our benchmark simulated models.  The left panel of Fig.~\ref{fig:ANGULAR} shows the radial distribution of flux for cosmic ray models with varying $\beta_p$.  Obviously, as $\beta_p$ increases the flux falls off less steeply with radius, but generically all of the cosmic ray models are relatively flat within 0.1 degrees and then fall-off steeply with increasing radius.  In comparison, the emission from the dark matter models (right panel of Fig.~\ref{fig:ANGULAR}) has a much flatter radial distribution, particularly when a significant substructure component is present.  With a smaller contribution from substructure, as is the case for DM\_HS, the gamma-ray flux falls steeply at small radii but eventually flattens around $R\sim0.3$ degrees where the contribution from substructure becomes significant.  For comparison, we also show in the right panel of Fig.~\ref{fig:ANGULAR} an alternative dark matter distribution to the centrally cuspy NFW model, the Burkert profile discussed in Sec.~\ref{sec:dm} (labeled Burk.~in the figure).  The radial profile of emission from the Burkert density distribution is very similar to our simulated NFW model at large radii, but flatter at small radii ($R<0.5$ degrees).  In practice, this difference can only be resolved for bright sources and at high energies where the PSF is smaller.

We now turn to our simulations and ask whether our simulated models can be detected as extended and whether a difference in spatial distribution can be observed for our cosmic ray versus dark matter models.  Unfortunately, a full joint fitting of the spectral and spatial distribution of a source cannot currently be carried out with the Fermi Science Tools, and here we choose to separate the spatial an spectral modeling.  For simplicity, we consider only one simulated spectrum (plus the two reference spatial models) each for the dark matter and cosmic ray cases, the high mass spectrum for dark matter and the soft spectrum for cosmic rays. We also fit only the higher S/N, 5 year simulations.  For each of the four simulations considered (DM\_HB, DM\_HS, CR\_SC, CR\_SF), we create images of the cluster plus extragalactic background for $E>100$ MeV and $R<3$ degrees binned to have 0.2 degree pixels.  This binning was chosen to give a reasonable number of counts per bin while still being smaller than the Fermi-LAT PSF at all but the highest energies.  While the PSF can be significantly reduced by only considering higher energies, we did not wish to reduce the photon counts from already faint sources.

First, we create a model of the energy averaged PSF for each spectral model considered by simulating a high flux point source with that spectrum.  An image of the point source was then created with the same binning, energy cut, and angular region as for our cluster simulations.  This image, re-normalized to one, is used as a PSF model, which is convolved with a given spatial model and then fit to the simulated cluster data.  All spatial fitting is done using the package \textit{Sherpa} \cite{sherpacit}, distributed as part of the Chandra data analysis software \textit{CIAO} \cite{ciaocit}.  We consider three different models for the spatial distribution as provided by \textit{Sherpa}: a delta function to test if a source is point-like, a Gaussian, and a $\beta$-model of the form
\begin{equation}
S(r) =  S_0 \left(1+ \left(\frac{r}{r_{core}}\right)^2\right)^{ - \alpha }.
\end{equation}
Neither a Gaussian nor a $\beta$-model exactly describes the spatial distributions shown in Fig.~\ref{fig:ANGULAR}, but these models give an indication of the extent and slope of the source distribution.  We fit using the maximum-likelihood based Cash statistic, CSTAT \cite{cash}, which is more appropriate than a $\chi^2$ statistic for data with few counts per bin but has the property that $\Delta C$ is distributed approximately as $\Delta \chi^2$ when the number of counts in each bin is $\gtrsim 5$ (as is the case for our images).  Our results are summarized in Tab.~\ref{tab:space5}.

\begin{table}
\hspace*{-2.cm}\begin{tabular}{c|c|c|cc|ccc}
\hline
Model & Bkgd? & Delta & Gaussian &  & $\beta$-Model & & \\
 & & Stat./DOF & Stat./DOF & FWHM (deg) & Stat./DOF & $\alpha$ &$r_{core}$ (deg) \\
\hline
DM\_HB & &\textbf{4518}/715 &\textbf{841.6}/714 &$3.26^{+0.06}_{-0.06}$ &767.8/713 &$0.74^{+0.08}_{-0.07}$ &$0.46^{+0.14}_{-0.12}$ \\
DM\_HB  &$\surd$ &\textbf{1784}/715 &\textbf{817.0}/714 &$3.44^{+0.14}_{-0.14}$ &\textbf{817.5}/713 &$10^{+?}_{-6.5}$ &$6.2^{+?}_{-2.9}$ \\
DM\_HS & &716.8/708 &510.2/707 &$3.36^{+0.38}_{-0.30}$ &493.6/706 &$0.43^{+0.13}_{-0.16}$ &$0.006^{+0.039}_{-0.006}$ \\
\hline
CR\_SC & &\textbf{855.0}/708 &785.7/707 &$0.46^{+0.06}_{-0.06}$ &780.5/706 &$2.0^{+0.6}_{-0.3}$ &$0.21^{+0.09}_{-0.05}$ \\
CR\_SC  &$\surd$ &\textbf{926.7}/715 &\textbf{859.2}/714 &$0.78^{+0.08}_{-0.08}$ &\textbf{805.4}/713 &$0.98^{+0.10}_{-0.10}$ &$0.06^{+0.03}_{-0.02}$ \\
CR\_SF & &\textbf{831.2}/708 &787.7/707 &$1.39^{+0.21}_{-0.20}$ &757.6/706 &$1.0^{+0.2}_{-0.1}$ &$0.10^{+0.09}_{-0.05}$ \\
\hline
\end{tabular}
\caption{ Results of spatial fitting to the 5 year simulations. Columns 3, 4, and 6 list the value of the Cash statistic for the fit over the number of degrees of freedom.  Values of the Cash statistic in bold indicate fits with a probability of less than 1\%.  Column 2 indicates whether or not the extragalactic diffuse background was included.  In row 2, for the $\beta$-model fit to DM\_HB including the gamma-ray background, $\alpha$ pegs at its upper limit and the upper limits on the parameters of the model are unconstrained. \label{tab:space5}}
\end{table}

Initially, we consider only the simulated cluster emission and neglect the extragalactic background.  Tab.~\ref{tab:space5} reveals a couple of trends.  First, none of the simulated clusters are consistent with a point source, with the exception of the very faint model DM\_HS whose spatial distribution is not well constrained.  Second, as expected, the dark matter models are more extended, in terms of the Gaussian FWHM, and have flatter profiles, in terms of the $\beta$-model slope $\alpha$ than the cosmic ray models.  Typically both the Gaussian and the $\beta$ models give acceptable fits to the data.  



Finally, we consider how well we can model the spatial distribution if the extragalactic background is included.  Here again we create images of our simulations with the same radius, binning, and energy range, but with the extragalactic diffuse emission included.  As in the spectral analysis, we use an outer annulus between 10 and 12 degrees to measure the background level.  The average background per pixel from this region is then added as a constant to our spatial modeling.  For the two fainter simulated clusters, DM\_HS and CR\_SF, unfortunately, the addition of the background means that the spatial model is not well constrained; the fits are poor and the errors on the model parameters are large.  For the brighter simulations, DM\_HB and CR\_SC, we again find a much better fit to an extended source than to a point source.  These fits are noted in Tab.~\ref{tab:space5}.  The fits typically worsen somewhat in terms of fit probability given the imperfect modeling of the background, but in general, the fit parameters such as the FWHM and slope are consistent within the errors with the no background results.

In summary, clusters are expected to be extended gamma-ray sources, and with the inclusion of the contribution from substructure to the dark matter distribution, the gamma-ray emission from dark matter annihilation is predicted to be flatter and more extended than for cosmic rays.  However, these differences can only be detected, given the extragalactic background, for bright sources with deep data.  If the extragalactic background used in this work is an overestimate of the Fermi background, due for example to AGN unresolved by EGRET, then the situation could improve significantly.  On the other hand, with real observations we will also have the additional uncertainty of imperfect knowledge of the spectrum.  As a final note, what we have presented here is a relatively simple approach to the spatial modeling of Fermi data based on tools currently available.  This analysis could be significantly improved through the development of tools to jointly fit the spectral and spatial distribution to arbitrary functions.

\subsection{Multi-wavelength counterparts}
\begin{figure}[t]
\begin{center}
\includegraphics[width=14.cm,clip]{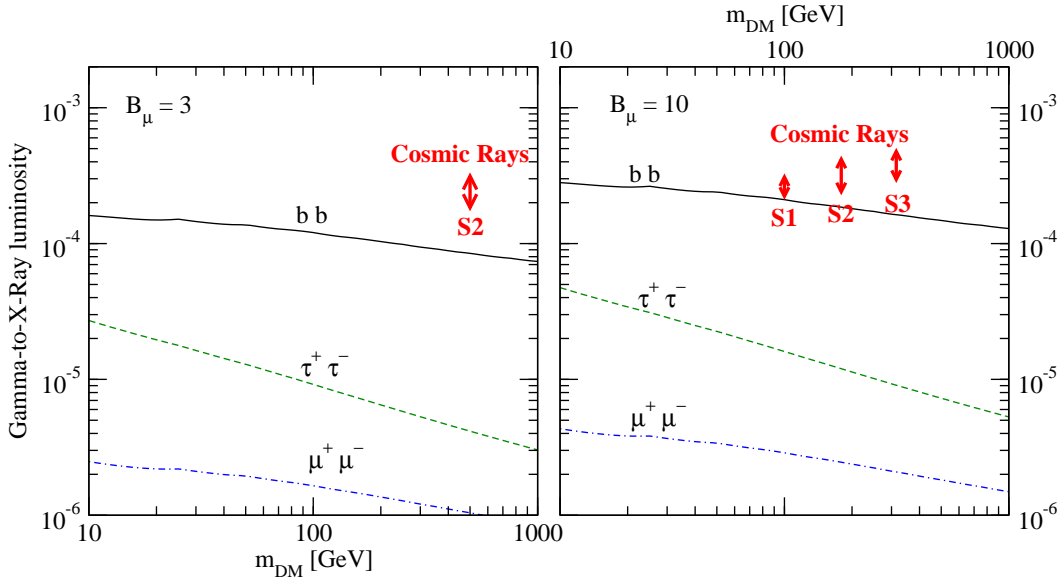}\\
\caption{The ratio of the gamma-ray to hard X-ray integrated flux for dark matter, as a function of the dark matter particle mass, for average magnetic fields $B=3\ \mu$G (left) and $B=10\ \mu$G (right). The red arrows indicate the range predicted for cosmic rays as assessed from the numerical simulations of \cite{pfrom}, with various assumptions on the cosmic ray physics.  \label{fig:xgamma}}
\end{center}
\end{figure}

Several recent studies highlighted the importance of secondary radiation
emitted from electrons and positrons produced in WIMP pair annihilation as
a powerful indirect dark matter diagnostic. The non-thermal population of
light stable leptons produced in dark matter halos via particle
annihilation radiates, in fact, through synchrotron, inverse Compton and
bremsstrahlung off the ICM gas, with peculiar spectral features. First
discussed in \cite{totani} for the case of galaxy clusters, the
multi-wavelength emission from dark matter annihilation was also studied
in detail in \cite{baltzwai} for galactic dark matter clumps and in
\cite{coladraco} for the dwarf spheroidal galaxy Draco.
Ref.~\cite{xrdwarf} extended those analyses and studied constraints on
particle dark matter properties from X-ray observations of nearby dwarf
galaxies. Other recent studies include an interpretation of the
significant non-thermal X-ray activity observed in the Ophiuchus cluster
in terms of IC scattering of dark-matter produced $e^+e^-$
\cite{ophiuchus}, an analysis of the broad-band dark-matter annihilation
spectrum expected from the Bullet cluster \cite{colabullet} and from the
galactic center region \cite{ullioregis}. In addition, radio emission from
$e^+e^-$ produced in dark matter annihilation was considered as a possible
source for the ``WMAP haze'' in the seminal paper of \cite{haze1}, and
subsequently analyzed in detail in \cite{haze2}, \cite{haze3} and
\cite{2008arXiv0801.4378H}. Other studies
\cite{gondolo,bertone0101134,aloisio0402588} have also addressed
synchrotron radiation induced by dark matter annihilation.

It was pointed out in \cite{ophiuchus} that a unique aspect of the
multi-wavelength spectrum from WIMP dark matter annihilation in clusters
is a strong emission at hard X-ray frequencies ($E_\gamma\gtrsim10$ keV),
with strikingly uniform spectral features, independent of the specific
WIMP model. Other studies, in particular \cite{xrdwarf}, showed that even
in systems where cosmic ray diffusion and leakage play an important role,
like dwarf galaxies, X-ray and gamma-ray constraints on particle dark
matter with current telescopes are comparable. It is therefore important
to assess whether information can be extracted by comparing the X-ray and
the gamma-ray emission from dark matter with the results of hydrodynamical
simulations of cosmic ray physics in clusters \cite{pfrom}. The latter
hard X-ray emission is thought to result from the inverse Compton
up-scattering of mostly cosmic microwave background photons by non-thermal
cosmic ray electrons accelerated to relativistic energies in the ICM
through the multiple mechanisms summarized in the Introduction.

While a dedicated analysis of the spectral features expected in the hard X-ray band for cosmic rays and for dark matter is beyond the scope of the present analysis, we wish here to compare the cumulative flux in these two frequency regions. Specifically, we consider the ratio of the integrated gamma-ray flux (for $E>0.1$ GeV)  to the integrated hard X-ray flux (for $E>10$ keV) as resulting from the numerical simulations of \cite{pfrom}, and quoted in Table 3 there. Fig.~\ref{fig:xgamma} indicates the results of \cite{pfrom} with red arrows. In the left panel we employ a scaling magnetic field value of 3 $\mu$G, while in the right panel of 10 $\mu$G, with the spatial distribution assumptions outlined in \cite{pfrom}. As specified in ref.~\cite{pfrom}, where we refer the Reader for further details, simulation S3 includes thermal shock heating, radiative cooling, star formation, Coulomb and hadronic cosmic ray losses, and cosmic rays from shocks and supernovae. Simulation S2 neglects the latter component, while S1 also neglects radiative cooling and star formation.

We also show in Fig.~\ref{fig:xgamma} the results of the same integrated hard X-ray and gamma-ray emission for dark matter models with a $b\bar b$, $\tau^+\tau^-$ and $\mu^+\mu^-$ dominant final state, as a function of the WIMP mass. While a $b\bar b$ final state, common to numerous supersymmetric dark matter models, produces a soft electron-positron spectrum, leptonic final states give rise to a harder emission. In turn, this corresponds to a larger IC flux in the hard X-ray band, and to a suppressed gamma-ray flux. To compute the multi-wavelength emission from WIMP annihilation, we use the same setup as the one outlined in \cite{ophiuchus}. Notice that the ratio of the gamma-ray to hard X-ray luminosity from WIMP annihilation is largely independent of the particular setup chosen for the dark matter profile and for substructures (see Sec.~\ref{sec:dm}).

We find that the generic expectation is for the ratio of gamma-ray to hard X-ray flux to be larger for cosmic rays and suppressed for dark matter. The suppression can be as large as a factor 100, for final states producing a hard electron-positron spectrum, such as the lightest Kaluza-Klein particle of Universal Extra Dimensions (UED) \cite{kkdm}. As we found in the spectral analysis and with the hardness ratio approach, the hardest cases to differentiate dark matter from cosmic rays are those with a very light dark matter particle. Comparing the left and the right panel of Fig.~\ref{fig:xgamma}, we also find that smaller values of the magnetic field enhance the difference in the gamma-to-hard X-ray luminosity for cosmic rays and for dark matter, while larger values tend to give a more blurry picture. While the assessment of the average intra-cluster magnetic field (via e.g. Faraday rotation measurements) might only give an order of magnitude estimate, we find that for expected values for the magnetic field and for dark matter particle models with electro-weak scale particles, the gamma-ray to hard X-ray ratio technique proposed here would give a rather robust handle. Notice that in theoretically favored dark matter particle models, such as supersymmetry, the ratio shown in fig.~\ref{fig:xgamma} would typically lie between the black solid line ($b\bar b$) and the green dashed line ($\tau^+\tau^-$). Other dark matter models, such as UED would have an even lower such ratio.

In short, we showed that the ratio of the integrated gamma-ray to hard X-ray flux in galaxy clusters can be used as a diagnostic for the discrimination of the origin of non-thermal phenomena, specifically astrophysical cosmic rays from dark matter annihilation. The generic expectation is that a dark matter induced signal would produce a brighter hard X-ray emission as opposed to cosmic rays, for a given detected gamma-ray flux.

\section{Gamma Ray Emission from Selected Galaxy Clusters and Groups}\label{sec:cat}
In this Section, we ask the question what are the best targets for the detection of gamma ray emission from clusters. We consider two large catalogs of galaxy clusters and groups, the HIGFLUCS Catalog \cite{higf} and a subset of the catalog produced by the GEMS project \cite{gems}. We describe below the assumptions we make to predict the gamma ray emission from cosmic rays and from dark matter for the objects considered in the two catalogs. In order to compare all objects, we make the same set of assumptions as far as both cosmic rays and dark matter are concerned: for instance, we assume the same ratio of gas to cosmic ray energy density ($X_p=0.01$, i.e. lower by a factor 10 than what we considered before for Coma) for all clusters and groups. In modeling the dark matter halo, we assume the same fraction of mass in substructures versus the host halo ($f_s=0.5$) and the resulting $\Delta^2=7\times 10^6$ as for the ``{\rm Boosted}''  setup (the fluxes corresponding to the ``Smooth'' case would have been roughly a factor 25 smaller, although this number depends on the cluster/group under consideration).

To compute the cosmic ray emission, the first step is to extract the electron scaling density $n_e$ from the X-ray data. We assume that the intra-cluster medium (ICM) density can be described by a beta model:
\begin{equation}
\rho(r)=\rho_0\left(1+\frac{r^2}{R_c^2}\right)^{-3\beta/2}.
\end{equation}
We follow here the approach outlined in \cite{Mohr:1999ya}, which assumes that the ICM is isothermal with temperature $T_X$, and we compute the scaling density of electrons $n_e$ as
\begin{equation}
n_e=\left(\frac{L_X\mu_H(1-3\beta)}{2\pi\mu_e\Lambda(T_X)R_c^3F(R_X)}\right)^{1/2},
\end{equation}
where
\begin{equation}
\mu_{e,p}=\frac{\rho(r)}{n_{e,H}(r)\ m_p},\quad {\rm with}\quad \mu_e\simeq1.167,\ \mu_H=1.400
\end{equation}
which corresponds to the assumption of fully ionized plasma with 30\% solar abundances \cite{feldman}, $L_X$ is the X-ray luminosity, $R_X$ is the X-ray detection radius, and the function $F$ is defined as \cite{Mohr:1999ya}
\begin{equation}
F(R)=\int_0^\infty\ {\rm d}s\Big[(1+s^2+(R/R_c)^2)^{1-3\beta}-(1+s^2)^{1-3\beta}\Big].
\end{equation}
For the radiative cooling coefficient $\Lambda(T)$ we assume the parametrization of \cite{Tozzi:2000cy}, i.e.
\begin{equation}
\Lambda(T)=C_1(kT)^\alpha+C_2(kT)^\beta+C_3,
\end{equation}
with
\begin{eqnarray}
\nonumber C_1=8.6\times 10^{-25}\ {\rm erg}\ {\rm cm}^3\ {\rm s}^{-1}\ {\rm keV}^{-\alpha}\\
\nonumber C_2=5.8\times 10^{-24}\ {\rm erg}\ {\rm cm}^3\ {\rm s}^{-1}\ {\rm keV}^{-\beta}\\
\nonumber C_3=6.3\times 10^{-24}\ {\rm erg}\ {\rm cm}^3\ {\rm s}^{-1}\\
\alpha=-1.7,\quad \beta=0.5.
\end{eqnarray}
We then use the cluster X-ray temperature $T_X$, the beta profile parameters $R_c$ and $\beta$, the cluster or group redshift and $R_{200}$ to compute the gamma-ray flux from cosmic rays. We assume an injection spectral index $\alpha_p=2.5$ (this parameter is unimportant here, as it just re-normalizes all the cosmic-ray induced gamma-ray fluxes but not the relative emission), and we normalize the cosmic ray energy density to a fraction $X_p=0.01$ of the gas energy density. For the cosmic ray source bias exponent $\beta_p$ we consider two opposite cases: a ``smooth'' case, where $n_{CR}\propto n_H$, hence $\beta_p=1$ (see Sec.~\ref{sec:cr}), and a ``cuspy'' case, where instead $\beta_p=-0.5$. We then integrate the gamma-ray emission above 0.1 GeV and out to $R_{200}$. Notice that this differs from the fluxes reported in Tab.~\ref{tab:spec}, where we integrated the gamma-ray flux over the entire angular region corresponding to our Fermi simulations ($10.5^\circ$). 

As far as the dark matter annihilation signal is concerned, we again use X-ray data to infer the dark matter density profile, assuming, as explained above, a Navarro-Frenk-White density distribution \cite{nfw}.
We assume the substructure setup outlined in Sec.~\ref{sec:dm} and in \cite{Colafrancesco:2005ji} (to which we refer the reader for further details), and the structure formation model of \cite{bullock}. Here, we assume again that the radial density distribution of substructure follows the same profile as Eq.~(\ref{eq:nfw}), but with a biased length scale $a^\prime\simeq7a$, see e.g. \cite{Nagai:2004ac, Diemand:2004kx}; we also assume an average bias in the concentration of subhalos versus host halos at equal mass $\langle c_s\rangle/\langle c_{\rm vir}\rangle=4$. Lastly, we assume that the fraction of mass in subhalos over the host halo mass $f_s=0.5$ (this setup corresponds to the Boosted or ``B'' case of sec.~\ref{sec:dm}). We then determine the scale radius $a$ and the scaling density $\rho_{0,\rm DM}$ in the following two ways. First, we take the $(R_{200},M_{200})$ and $(R_{500},M_{500})$ pairs as determined from X-ray data, and solve for $a$ and  $\rho_{0,\rm DM}$. Alternatively, we use the model of \cite{bullock} to relate the concentration of a given cluster to its mass and redshift (in practice for our low-redshift samples the latter does not matter significantly), and only make use of one of the $(R,M)$ pairs from X-ray data. Remarkably, we find that the two procedures yield very similar results in the final dark matter induced gamma-ray flux, to the level of better than 10\%.

On the particle dark matter side, we assume a particle mass of 40 GeV, a pair annihilation cross section $\langle\sigma v\rangle=6\times 10^{-26}\ {\rm cm}^3/{\rm s}$ and 23 gamma rays per dark matter annihilation event above 0.1 GeV, as Monte Carlo simulations indicate is the case for a $b\bar b$ annihilation final state, ubiquitous e.g. in supersymmetric models \cite{dmreviews}. This particle DM model corresponds to the low mass (L) model of Sec.~\ref{sec:dm}. Once the DM setup is specified, the gamma ray flux from dark matter annihilations is then simply the integral over the line of sight of the number density squared of dark matter particles, times the annihilation rate. The resulting fluxes we report, $\phi_0$, can easily be rescaled for other masses, pair annihilation cross sections and annihilation final states:
\begin{equation}
\phi^\prime=\phi_0\left(\frac{40\ {\rm GeV}}{m_{\rm WIMP}}\right)^2\left(\frac{\langle\sigma v\rangle}{6\times 10^{-26}\ {\rm cm}^3/{\rm s}}\right)\left(\frac{N_\gamma^{E>0.1\ {\rm GeV}}}{23}\right).
\end{equation}
In the case of the GEMS catalog  \cite{gems}, the group masses were not indicated. We need however that information to reconstruct the dark matter density profile. We thus assume that the ICM is in hydrostatic equilibrium and isothermal, and set \cite{higf}
\begin{equation}
M(<r)=\frac{3kT_X\ r^3\beta}{\mu m_p G_N}\left(\frac{1}{r^2+R_c^2}\right)
\end{equation}
where $m_p$ is the proton mass, $\mu\simeq0.61$ is the mean molecular weight and $G_N$ is the gravitational constant.

\subsection{Gamma-ray Emission from Clusters: the HIGFLUCS Catalog}
\begin{figure}[t]
\begin{center}
\mbox{\hspace*{-1cm}\includegraphics[width=18.5cm,clip]{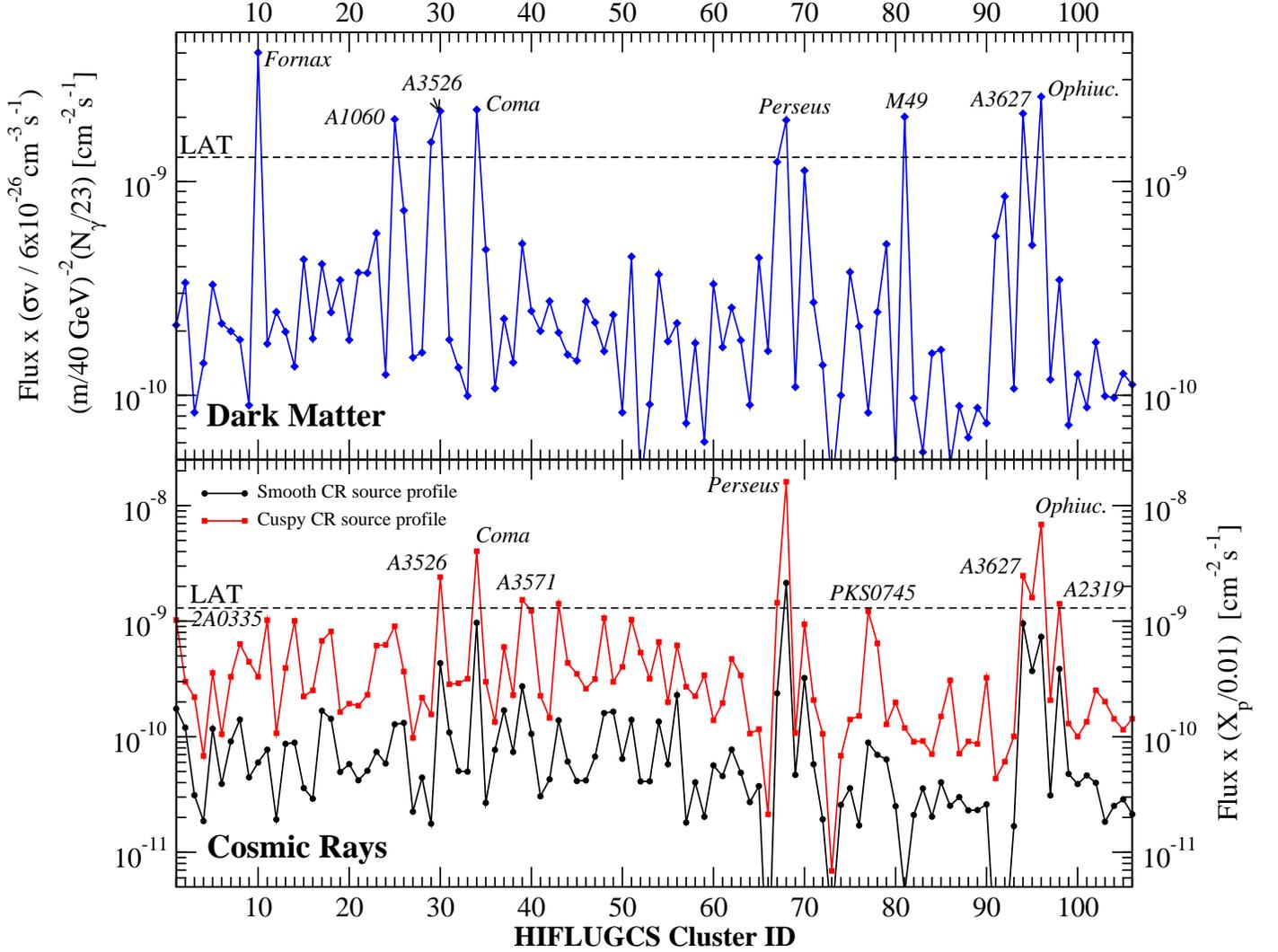}}\\
\caption{Upper panel: predicted gamma-ray flux from dark matter annihilation in the extended HIGFLUCS catalog, with the assumptions for the dark matter profile specified in the text, and for a particle mass of 40 GeV, a $b\bar b$ dominant annihilation final state and a pair annihilation cross section of $6\times 10^{-26}\ {\rm cm}^3/{\rm s}$, Lower panel: X-ray data-driven predictions for the gamma-ray emission from cosmic rays in clusters and groups in the extended HIGFLUCS catalog. The red line refers to a cuspy spatial distribution for the cosmic ray sources ($\beta_p=-0.5$), while the black line to a smooth distribution ($\beta_p=1$).\label{fig:clusters1}}
\end{center}
\end{figure}
We applied the procedure described above to the HIGFLUCS Catalog, including the clusters and groups from the Extended Sample \cite{higf}. The HIGFLUCS Catalog includes candidates from several input catalogs, and it includes 63 clusters featuring an X-ray flux in the 0.1-2.4 keV range larger than $2.0\times10^{-11}\ {\rm ergs}\ {\rm s}^{-1}\ {\rm cm}^{-2}$, with galactic latitude $b>20.0^\circ$ and outside two excluded areas towards the Magellanic clouds and the Virgo cluster. In addition, we also include the Extended Sample, with 43 more clusters, bringing the total number of clusters to 106. We collect the flux ID numbers, names, predicted gamma-ray fluxes, ranking and ratio of cosmic ray to dark matter signal in Table \ref{tab:hf1}, \ref{tab:hf2} and \ref{tab:hfext} in the Appendix.

We show the flux of gamma rays from dark matter annihilation in the upper panel of Fig.~\ref{fig:clusters1}. We find that the eight clusters with the expected largest dark matter induced gamma ray flux are the nearby large clusters Perseus, Coma, Ophiuchus and Abell 1060, 3526 and 3627. In addition the two groups M49 and Fornax also have very large fluxes, the latter in particular giving the largest one in the entire sample. The scatter in flux over the HIGFLUCS Catalog ranges over more than two orders of magnitude, but for our nominal choices for the dark matter particle properties and density distribution the typical flux from dark matter from low-redshift clusters lies typically between $10^{-10}$ and $10^{-9}\ {\rm cm}^{-2}\ {\rm s}^{-1}$.

The lower panel of Fig.~\ref{fig:clusters1} shows our results for the cosmic-ray induced gamma-ray flux. We show the results for both a smooth ($\beta_p=1$) and a cuspy ($\beta_p=-0.5$) primary cosmic ray source distribution. The cuspy source distribution typically boosts the flux by around one order of magnitude, and is expected e.g. in clusters with bright active galactic nuclei. The actual cosmic ray flux is expected to be somewhere in between the red and the black line. The clusters expected to be brightest in gamma-rays from cosmic ray interactions include Perseus, Coma, Ophiuchus, Abell 3526, 3571, 3627 and 2319, as well as the bright cooling flow clusters 2A0335 and PKS0745. 

We remark that although our data-driven analytic approach is different and new with respect to other attempts at predicting which clusters are brighter in gamma rays, we substantially agree with previous analyses, including \cite{reimer} which uses the ratio of the cluster mass over distance squared, and \cite{pfrom}, which makes use of the results of numerical simulations to assess scaling relations that then are used to predict the gamma-ray emission. We naturally agree with \cite{nagaiando}, which uses a similar approach to ours.  We note that these previous analyses predict only the gamma ray flux from cosmic rays and neglect dark matter annihilation.

\begin{figure}[t]
\begin{center}
\includegraphics[width=15.cm,clip]{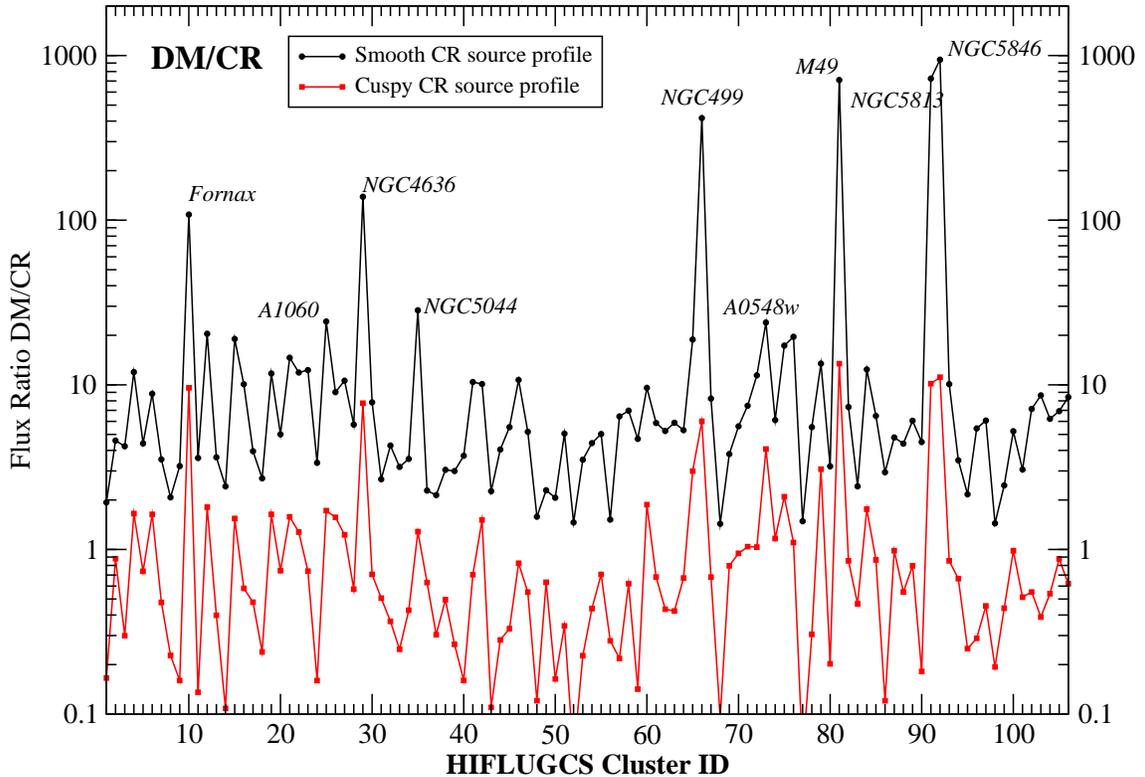}\\
\caption{The ratio of the gamma-ray emission from dark matter and from cosmic rays for the clusters and groups in the extended HIGFLUCS catalog. The red and black lines refer to a cuspy and a smooth cosmic ray source distribution, as in Fig.~\ref{fig:clusters1}. \label{fig:dmcr}}
\end{center}
\end{figure}
Fig.~\ref{fig:dmcr} shows the ratio of the dark matter to cosmic ray gamma-ray flux for the extended HIGFLUCS sample. As in Fig.~\ref{fig:clusters1} we use the black line to indicate fluxes corresponding to a smooth cosmic ray source profile and a red line for the cuspy profile. In either approach, we notice several outliers, featuring a very high dark matter induced emission compared to the cosmic ray contribution. Candidates include NGC 5846, 5813, 499, 5044 and 4636, as well as Abell 1060, 0548w, Fornax and M49. Interestingly, all these candidates are nearby poor clusters or galaxy groups. Even more interestingly, some of them (like M49 and Fornax) feature some of the largest predicted gamma-ray emissions from dark matter annihilation, making them particularly interesting candidates for Fermi observations. Even with extreme assumptions on the cosmic ray emission ($\beta_p=-0.5$), we find that the dark matter signal could be a factor 10 larger than that of the cosmic ray induced flux. Searches for gamma-ray signals from these promising set of objects will be extremely valuable even with only one year of Fermi data.

\begin{figure}[t]
\begin{center}
\includegraphics[width=14.cm,clip]{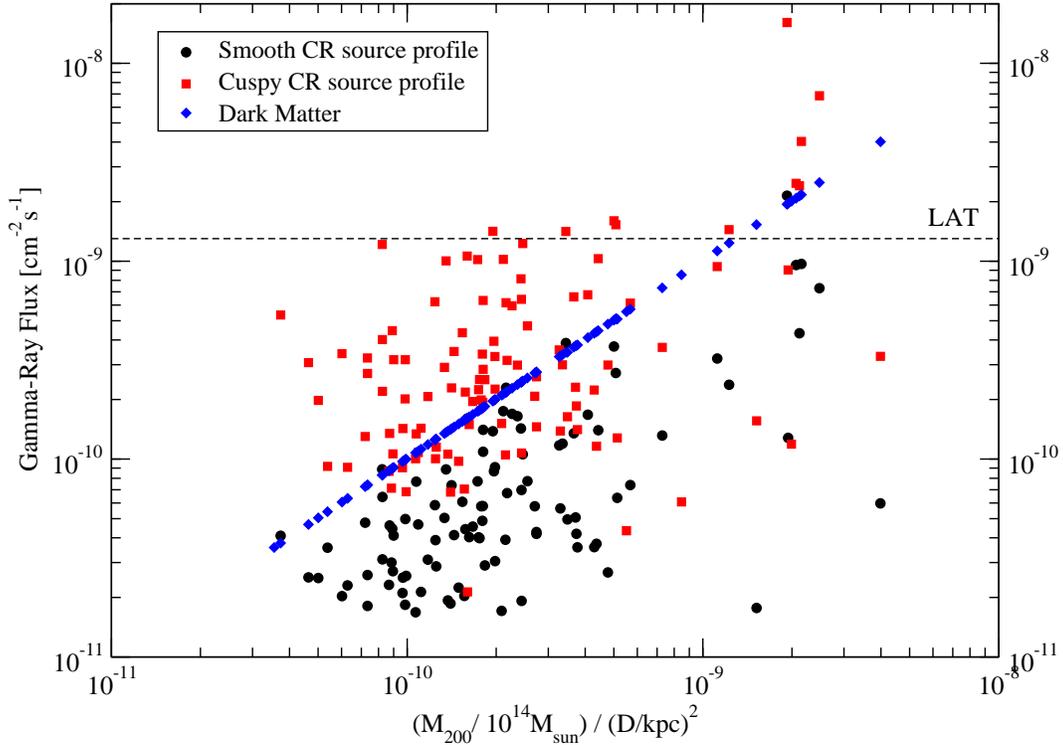}\\
\caption{Correlation between the gamma-ray fluxes (from cosmic rays and dark matter) and the ratio of mass to distance squared, for objects in the HIGFLUCS catalog. \label{fig:corr_MD2}}
\end{center}
\end{figure}
We show in Fig.~\ref{fig:corr_MD2} the correlation between the cluster mass to distance squared ratio and the predicted gamma-ray flux from cosmic rays and dark matter. We find a very tight correlation for the dark matter emission, which mainly depends on what we assumed for the reconstruction of the dark matter density profiles. On the other hand, the scatter in the beta model parameters induce a significant scatter for the cosmic ray induced gamma-ray emission, with significant outliers both in excess and in deficit.

\begin{figure}[t]
\begin{center}
\includegraphics[width=14.cm,clip]{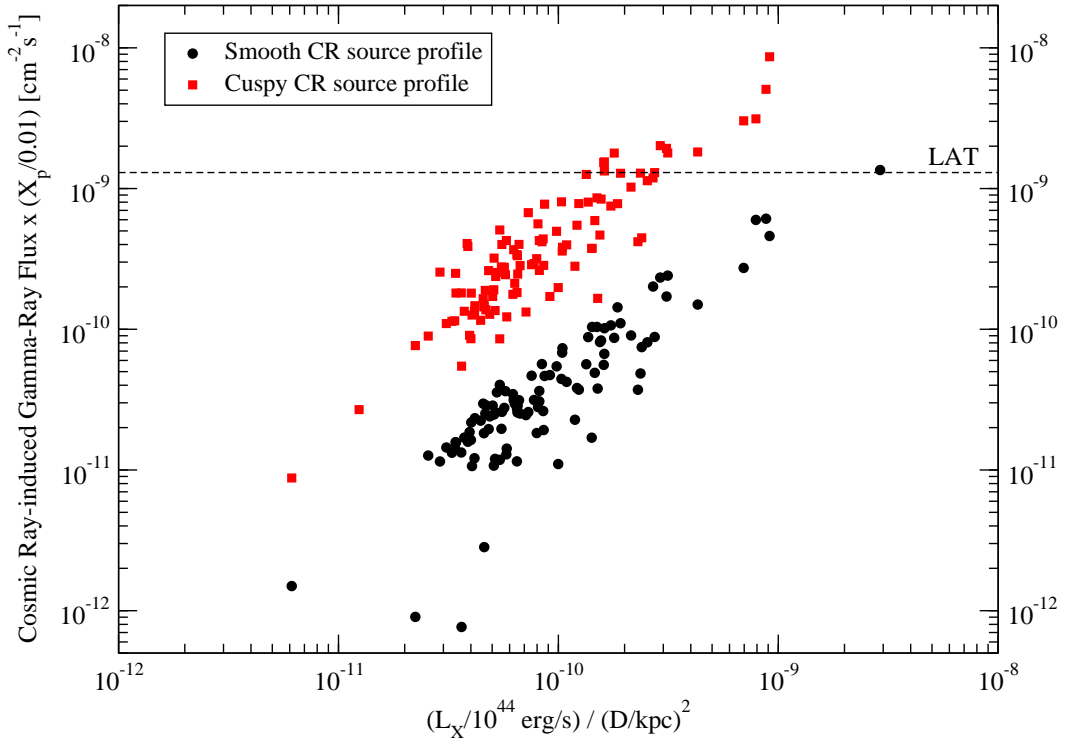}\\
\caption{Correlation between the gamma-ray fluxes from cosmic rays and the ratio of X-ray luminosity to distance squared, for objects in the HIGFLUCS catalog. \label{fig:corr_LXD2}}
\end{center}
\end{figure}
Fig.~\ref{fig:corr_LXD2} shows a correlation between the predicted gamma-ray luminosity from cosmic rays and the X-ray luminosity over distance squared for clusters in the extended HIGFLUCS sample. Although there are a few outliers, the correlation is rather tight, and close to linear. The dark matter induced emission, instead, does not show any significant correlation with the X-ray luminosity. Also, we do not find any other strong correlations between the predicted gamma-ray fluxes and X-ray related quantities.

As a side comment, we remark that two of the three clusters for which \cite{Wolfe:2008qm} tentatively associates an unidentified EGRET source with radio sources in the NRAO VLA and Westerbork Northern sky survey catalogs appear also in our list, namely Abell 85 and 1914. While we predict Abell 85 be quite luminous in cosmic-rays-induced gamma-ray emission (ranking 12th to 16th out of 130, depending on the cosmic ray model), Abell 1914 is not predicted to be particularly bright. In addition, both clusters have a low dark matter to cosmic ray gamma-ray luminosity ratio, ranking respectively 124th and 87th. A dark matter interpretation for the tentative gamma-ray emission from these clusters seems therefore disfavored.

\subsection{Gamma-ray Emission from Groups: the GEMS Catalog}
\begin{figure}[t]
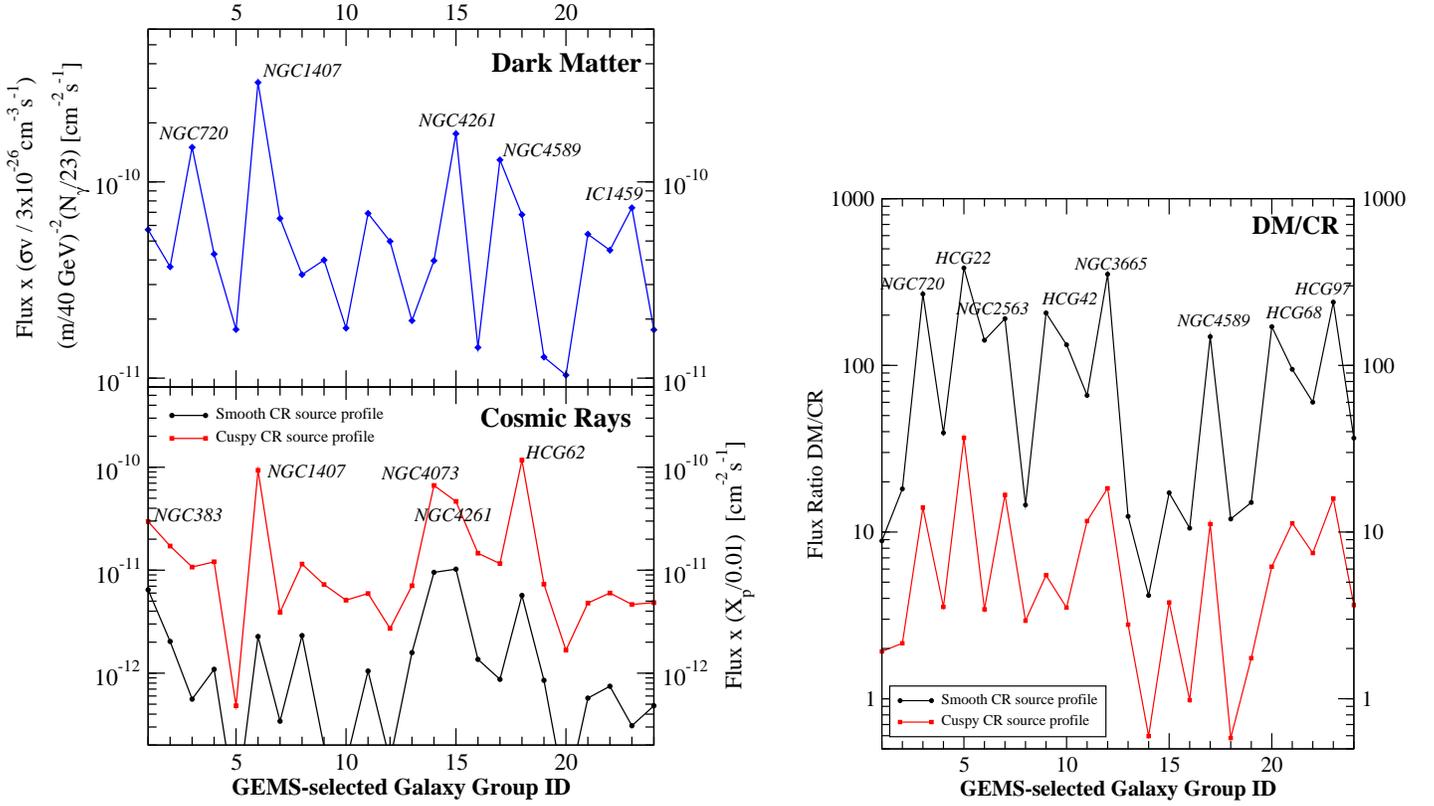

\begin{center}
\mbox{\hspace*{-1cm}\includegraphics[width=10.cm,clip]{groups_v2.eps}\qquad \includegraphics[width=8.cm,clip]{dmcr_groups_v2.eps}}\\
\caption{Gamma-ray fluxes from dark matter and cosmic rays (left) and their ratio (right) for our sub-sample of groups in the GEMS catalog. \label{fig:groups}}
\end{center}
\end{figure}
Given the results of the previous Section, we decided to consider an alternate sample of nearby groups to (1) check whether our predictions with the HIGFLUCS sample depend on the X-ray data analysis performed in \cite{higf} and (2) identify other potentially promising candidates for gamma ray searches for particle dark matter. We thus considered the catalog provided by the GEMS project \cite{gems}. We excluded from our sample the following:
\begin{enumerate}
\item groups with an X-ray flux less than 3$\sigma$ above the background level (U-sample \cite{gems})
\item groups where the detectable extent of group emission is $r_{\rm ext}<60$ kpc, i.e. where the emission appears to be associated with the halo of an individual galaxy instead of being genuinely intergalactic (H-sample \cite{gems})
\item groups for which some of the relevant X-ray information was not available
\item groups already included in the HIGFLUCS sample
\end{enumerate}
After the cuts described above, the GEMS sub-sample we used was limited to 24 groups. In passing, we notice that for the groups in the last item we obtained a remarkable agreement between the predictions for the gamma-ray fluxes we obtained with the HIGFLUCS and with the GEMS data.

We show in Fig.~\ref{fig:groups} our results for the dark matter and cosmic ray induced gamma-ray fluxes (left) and the dark matter to cosmic ray ratio (right), with the same conventions as we used in Fig.~\ref{fig:clusters1} and \ref{fig:dmcr}. We find that the gamma-ray emission associated to the groups in the GEMS sample is generically rather faint, with only one group (NGC1407) predicted to have a dark matter signal above $2\times10^{-10}\ {\rm cm}^{-2}\ {\rm s}^{-1}$ and only three groups (HCG62, NGC4073 and NGC1407) that might have a gamma-ray emission from cosmic rays at the level of $\sim10^{-10}\ {\rm cm}^{-2}\ {\rm s}^{-1}$. More importantly, though, we confirm with the GEMS sample that nearby groups have a potentially very large ratio of dark matter to cosmic ray gamma-ray emission. We find in fact several candidates with a $\sim100$ times larger dark matter than cosmic ray gamma-ray emission, assuming a smooth cosmic ray source profile (see the right panel of Fig.~\ref{fig:groups}).

\section{Discussion, Summary and Conclusions}\label{sec:concl}

The detection of gamma rays from clusters of galaxies might be a milestone in the scientific legacy of the Fermi Gamma-ray Space Telescope. Clusters are known to host a variety of high-energy phenomena that could fuel cosmic rays, producing, in turn, gamma rays as a result of collisions with the ICM gas. Being the largest bound dark matter structures, it is also reasonable to expect that clusters  feature a significant gamma-ray emission from the annihilation of dark matter particles. The scope of the present theoretical study was to investigate how to tell apart these two potential mechanisms of gamma-ray production with data from Fermi.  

One potential source of gamma-rays from cluster regions that we have not considered here are the bright, central AGN known to be present in some clusters (NGC 1275 in Perseus and M87 in Virgo), which could inhibit our ability to detect faint extended gamma-ray emission.  However, this source of contamination is not expected from some of the best candidate, nearby clusters, like the Coma cluster, where strong radio galaxies are not observed.  The best method to detect diffuse gamma-ray emission, from cosmic rays or dark matter annihilation, will be to concentrate on those clusters lacking bright AGN.  For clusters with central AGN, the expected extended nature of the cluster gamma-ray emission may reveal this component, and the well-known multiwavelength (radio and X-ray) AGN spectra will allow modeling of the contribution of these sources to the gamma-ray emission.

By making simple analytical assumptions on the cosmic ray spectra and source distribution and on the dark matter particle properties and density distribution, we proposed a set of various benchmark models for both gamma-ray production mechanisms under investigation. We believe the set of models we considered is representative of the variety of possibilities one can realistically expect to encounter in clusters of galaxies. We then proceeded to simulate the expected 1 and 5 years gamma-ray signals with the Fermi Science Tools, and analyzed our results.

We summarize below the main results of the present theoretical study.
\begin{itemize}
\item We find that for cosmic rays the absolute signal to noise peaks at around one degree, while for dark matter it can peak at larger regions of interest, possibly as large as 3 degrees. Since the Fermi instrument response function forces us to consider relatively large angular regions to include low energy photons, we conclude that the best region of interest for gamma-ray studies of cluster of galaxies is around 3 degrees.
\item The spectral analysis of the simulated signal for our benchmark models shows that gamma ray emission from dark matter annihilation with a relatively large dark matter particle mass ($m_{{\rm WIMP}}>50$ GeV) can be distinguished from a cosmic ray spectrum even for fairly faint sources.  Distinguishing a cosmic ray spectrum from a low dark matter particle mass appears to be more challenging, and would require deep data and/or a bright source.
\item We defined optimal energy bands for a simple hardness ratio, (H -- S)/(H + S), estimate of the spectrum.  The energy bands we propose are S $= 0.15 \leq E < 0.7$ GeV and H $= E \geq 0.7$ GeV. The hardness ratio is correlated to the nature of the emission and is predicted to be negative for a cosmic ray emission and positive for a dark matter annihilation signal.  Similar to the full spectral analysis, the hardness ratios for a simulated low mass dark matter particle model and a hard cosmic ray model are similar within the statisitcal errors. However, it is possible to use this simple two energy band ratio to distinguish cosmic ray and high mass dark matter particle models.
\item The level of the gamma-ray emission from clusters produced by cosmic rays can be comparable to that from dark matter annihilation. In the case of a mix of the two emissions, we find that for bright enough sources, if the dark matter contribution to the cluster flux is significant and the particle mass is not very low ($m_{WIMP} > 40$ GeV) the presence of a dark matter component can be seen even in the presence of a significant gamma-ray flux from cosmic rays.  However, tight constraints on the model parameters ($m_{WIMP}$ and $\alpha_p$) may be problematic unless the dark matter particle mass is known.
\item Our cluster gamma-ray simulated emissions appear, after data reduction, as extended rather than point sources, with extensions which depend on the spatial models and on the emission mechanism.  However, determining the spatial distribution with Fermi will be a challenging task requiring an optimal control of the backgrounds.
\item We showed that the ratio of the integrated gamma-ray to hard X-ray flux in galaxy clusters can be used as a diagnostics for the discrimination of the origin of non-thermal phenomena. The generic expectation is that a dark matter induced signal would produce a brighter hard X-ray emission as opposed to cosmic rays, for a given gamma-ray flux.
\item We presented X-ray data-driven predictions for the gamma-ray flux from 130 clusters and groups of galaxies in the HIGFLUCS and GEMS catalogs. We found that the clusters with the brightest gamma-ray emission from cosmic rays include the Perseus, Coma, Ophiuchus, Abell 3627 and Abell 3526 clusters; the most luminous clusters in dark matter emission are predicted to be the Fornax group, Ophiuchus, Coma, Abell 3526 and Abell 3627.
\item We discovered that the objects with the largest dark matter to cosmic ray gamma-ray luminosity in our sample are groups and poor clusters. In particular, the highest ratios are associated to the groups NGC 5846, 5813 and 499, M49 and HCG 22. Of these, M49 ranks overall 6th/130 in terms of predicted dark matter induced gamma-ray emission, NGC 5846 ranks 12th and NGC 5813 15th; Fornax also has a relatively high dark matter to cosmic ray gamma-ray flux, and is the brightest object in dark matter emission. All these objects are very promising candidates for a search for gamma-ray emission with the Fermi telescope that could potentially be related to particle dark matter: specifically, as shown in the Appendix, we predict, for the DM\_HS setup, an integrated flux of photons above 0.1 GeV of $4\times 10^{-9}\ {\rm cm}^{-2}{\rm s}^{-1}$ for Fornax and of  $2\times 10^{-9}\ {\rm cm}^{-2}{\rm s}^{-1}$ for M49, both around the projected Fermi point-source sensitivity for one year of data. With the same setup, we predict a flux of $0.8\times 10^{-9}\ {\rm cm}^{-2}{\rm s}^{-1}$ and $0.5\times 10^{-9}\ {\rm cm}^{-2}{\rm s}^{-1}$ for NGC 5846 and 5813, respectively, which would make them detectable or marginally detectable sources for Fermi within the anticipated mission lifetime.
\end{itemize}



  
\section*{Acknowledgments}
We acknowledge useful discussions and inputs from several members of the Fermi Collaboration, in particular James Chiang, Johann Cohen-Tanugi, Jan Conrad, Joakim Edsjo, Robert Johnson, Greg Madejski, Aldo Morselli, Eric Nuss, Troy Porter, Joel Primack, Olaf Reimer, and others involved in the New Physics/Dark Matter and in the AGN/Clusters Fermi Science Working Groups. T.E.J. is grateful for support from the Alexander F. Morrison Fellowship, administered through the University of California Observatories and the Regents of the University of California. S.P. is partly supported by US DoE Contract DEFG02-
04ER41268, NASA Grant Number NNX08AV72G and NSF Grant PHY-0757911.



\clearpage
\section*{Appendix}

We report here the X-ray data driven gamma-ray flux predictions for the HIGFLUCS catalog (tab.~\ref{tab:hf1} and \ref{tab:hf2}, plus members of the extended catalog \ref{tab:hfext}) and for the GEMS catalog (tab.~\ref{tab:gems}). The first column indicates the object ID used in our plots, the second one specifies the object name. The third column shows the predicted gamma-ray flux from dark matter annihilation for the model setup DM\_LB of sec.~\ref{sec:dm} (featuring a particle mass of 40 GeV, a pair annihilation cross section of $6\times 10^{-26}\ {\rm cm}^3/{\rm s}$, and 23 photons with an energy above 0.1 GeV per annihilation), and the fourth column the overall ranking (considering at the same time the extended HIGFLUCS catalog and the GEMS catalog). The next columns indicate the predictions for a cosmic ray origin, under the assumption of a smooth and of a cuspy source distribution, as well well as the overal ranking according to the corresponding emission. Lastly, the rightmost columns indicate the ratio of the dark matter to cosmic ray gamma-ray emission, with ranking.

The flux predictions quoted here can be easily rescaled for different particle DM models as:
\begin{equation}
\phi^\prime=\phi_0\left(\frac{40\ {\rm GeV}}{m_{\rm WIMP}}\right)^2\left(\frac{\langle\sigma v\rangle}{6\times 10^{-26}\ {\rm cm}^3/{\rm s}}\right)\left(\frac{N_\gamma^{E>0.1\ {\rm GeV}}}{23}\right),
\end{equation}
where $\phi_0$ is the integrated gamma-ray flux above 0.1 GeV we quote in the following tables and $\phi^\prime$ is the corresponding re-scaled flux. Similarly, for the cosmic-ray predictions, the scaling with the fraction of energy density in cosmic rays $X_p$ (which we here set at 1\%) is simply
\begin{equation}
\phi^\prime=\phi_0\left(\frac{X_p}{0.01}\right).
\end{equation}

In the tables, we mark in red the top-ten objects in every category we consider (gamma-ray emission from DM annihilation, emission from cosmic-ray inelastic interactions, and ratio of the DM-to-cosmic-rays projected emission).

In the figures, we also indicate with a horizontal dashed line an estimate of the high-latitude point source sensitivity of Fermi-LAT over 5 years ($\sim1.3\times 10^{-9}$ photons per ${\rm cm}^2$ per s). We predict, with the assumptions we made here for the DM setup, that on the order of 10 clusters will produce a large enough DM annihilation emission of gamma rays. Over the anticipated Fermi lifetime of 10 years, 5 or so more clusters might also be detectable according to the present predictions.
\\[0.2cm]

\begin{table}[!h]
\begin{tabular}{rlrrrrrrrr}
\hline
ID & Name & DM\_LB & DM\_LB,& CR\_S & CR\_S, & CR\_C & CR\_C,& DM\_LB/CR\_S & DM\_LB/CR\_S, \\
 & & & rank & & rank & & rank & & rank\\
\hline
   1 & A0085    &  213.1 &  47 &  110.3 &  12 &  1285.0 &  16 &    1.9 & 124 \\
  2 & A0119    &  336.4 &  30 &   73.3 &  25 &   382.4 &  47 &    4.6 &  86 \\
  3 & A0133    &   83.1 &  98 &   19.6 &  74 &   277.3 &  61 &    4.2 &  92 \\
  4 & NGC507   &  141.3 &  74 &   11.8 &  96 &    85.5 & 104 &   11.9 &  40 \\
  5 & A0262    &  329.2 &  32 &   74.7 &  24 &   446.4 &  36 &    4.4 &  89 \\
  6 & A0400    &  216.6 &  46 &   24.5 &  68 &   132.6 &  91 &    8.8 &  54 \\
  7 & A0399    &  199.5 &  50 &   56.6 &  28 &   418.2 &  41 &    3.5 & 101 \\
  8 & A0401    &  182.2 &  54 &   87.9 &  19 &   801.0 &  24 &    2.1 & 122 \\
  9 & A3112    &   89.8 &  94 &   28.0 &  57 &   561.3 &  31 &    3.2 & 105 \\
 10 & FORNAX   &  {\color{red}\bf4016.0} &    {\color{red}\bf1} &   37.2 &  43 &   419.0 &  40 &  108.1 &  16 \\
 11 & 2A0335   &  174.4 &  62 &   48.5 &  33 &  1286.0 &  15 &    3.6 &  99 \\
 12 & IIIZw54  &  245.3 &  40 &   12.0 &  95 &   135.8 &  88 &   20.4 &  25 \\
 13 & A3158    &  198.1 &  51 &   54.5 &  31 &   496.8 &  34 &    3.6 &  98 \\
 14 & A0478    &  136.5 &  76 &   56.4 &  29 &  1259.0 &  17 &    2.4 & 116 \\
 15 & NGC1550  &  432.4 &  22 &   22.8 &  71 &   280.4 &  60 &   19.0 &  27 \\
 16 & EXO0422  &  184.5 &  53 &   18.3 &  78 &   317.6 &  55 &   10.1 &  50 \\
 17 & A3266    &  410.8 &  23 &  103.9 &  14 &   858.9 &  21 &    4.0 &  95 \\
 18 & A0496    &  244.5 &  41 &   90.4 &  17 &  1024.0 &  20 &    2.7 & 112 \\
 19 & A3376    &  346.3 &  29 &   29.6 &  53 &   212.0 &  71 &   11.7 &  42 \\
 20 & A3391    &  181.7 &  56 &   36.3 &  45 &   244.2 &  69 &    5.0 &  83 \\
 21 & A3395s   &  375.0 &  25 &   25.7 &  63 &   237.3 &  70 &   14.6 &  33 \\
 22 & A0576    &  373.6 &  26 &   31.4 &  49 &   293.0 &  56 &   11.9 &  41 \\
 23 & A0754    &  572.4 &  14 &   46.6 &  36 &   774.2 &  27 &   12.3 &  38 \\
 24 & HYDRA-A  &  125.1 &  81 &   37.2 &  42 &   782.2 &  25 &    3.4 & 104 \\
 25 & A1060    &  {\color{red}\bf1958.0} &    {\color{red}\bf7} &   80.8 &  23 &  1137.0 &  19 &   24.2 &  23 \\
 26 & A1367    &  732.9 &  13 &   81.2 &  22 &   467.1 &  35 &    9.0 &  52 \\
 27 & MKW4     &  150.4 &  70 &   14.2 &  89 &   122.5 &  94 &   10.6 &  45 \\
 28 & ZwCl1215 &  158.5 &  67 &   27.6 &  58 &   275.9 &  62 &    5.7 &  72 \\
 29 & NGC4636  &  {\color{red}\bf1530.0} &    {\color{red}\bf9} &   11.0 &  99 &   197.7 &  72 &  138.8 &  14 \\
 30 & A3526    &  {\color{red}\bf2137.0} &    {\color{red}\bf4} &   {\color{red}\bf273.0} &  {\color{red}\bf5} &   {\color{red}\bf3028.0} &    {\color{red}\bf5} &    7.8 &  58 \\
\hline
\end{tabular}
\caption{Gamma-ray emission predictions for the groups and clusters in the HIGFLUCS catalog (first part). Fluxes are in units of $10^{-12}{\rm ph}/({\rm cm}^{2}\ {\rm s})$\label{tab:hf1}}
\end{table}
\begin{table}
\begin{tabular}{rlrrrrrrrr}
\hline
ID & Name & DM\_LB & DM\_LB,& CR\_S & CR\_S, & CR\_C & CR\_C,& DM\_LB/CR\_S & DM\_LB/CR\_S, \\
 & & & rank & & rank & & rank & & rank\\
\hline
31 & A1644    &  182.1 &  55 &   68.1 &  26 &   359.5 &  51 &    2.7 & 113 \\
 32 & A1650    &  134.7 &  77 &   31.5 &  48 &   367.7 &  50 &    4.3 &  91 \\
 33 & A1651    &   99.3 &  88 &   31.3 &  50 &   400.9 &  43 &    3.2 & 107 \\
 34 & COMA     & {\color{red}\bf2170.0} &   {\color{red}\bf3} &  {\color{red}\bf610.8} &   {\color{red}\bf2} &  {\color{red}\bf5073.0} &   {\color{red}\bf3} &    3.6 & 100 \\
 35 & NGC5044  &  481.1 &  19 &   17.0 &  81 &   375.1 &  49 &   28.4 &  22 \\
 36 & A1736    &  107.8 &  85 &   47.3 &  34 &   171.0 &  81 &    2.3 & 118 \\
 37 & A3558    &  228.1 &  43 &  106.6 &  13 &   750.2 &  28 &    2.1 & 121 \\
 38 & A3562    &  142.5 &  73 &   46.8 &  35 &   287.5 &  57 &    3.0 & 109 \\
 39 & A3571    &  512.0 &  16 &  {\color{red}\bf171.0} &   {\color{red}\bf9} &  {\color{red}\bf1929.0} &   {\color{red}\bf7} &    3.0 & 110 \\
 40 & A1795    &  247.9 &  38 &   66.8 &  27 &  1546.0 &  11 &    3.7 &  97 \\
 41 & A3581    &  199.9 &  49 &   19.2 &  76 &   284.4 &  58 &   10.4 &  47 \\
 42 & MKW8     &  275.6 &  34 &   27.2 &  59 &   182.4 &  75 &   10.1 &  48 \\
 43 & A2029    &  196.4 &  52 &   86.9 &  20 &  {\color{red}\bf1787.0} &  {\color{red}\bf10} &    2.3 & 119 \\
 44 & A2052    &  154.9 &  69 &   38.3 &  40 &   548.6 &  32 &    4.0 &  94 \\
 45 & MKW3S    &  145.2 &  72 &   26.2 &  60 &   438.6 &  37 &    5.5 &  75 \\
 46 & A2065    &  274.9 &  35 &   25.6 &  64 &   333.0 &  53 &   10.7 &  44 \\
 47 & A2063    &  219.3 &  44 &   42.3 &  38 &   398.1 &  45 &    5.2 &  80 \\
 48 & A2142    &  160.8 &  66 &  101.7 &  16 &  1333.0 &  13 &    1.6 & 125 \\
 49 & A2147    &  237.8 &  42 &  103.8 &  15 &   375.9 &  48 &    2.3 & 117 \\
 50 & A2163    &   83.1 &  99 &   40.3 &  39 &   508.0 &  33 &    2.1 & 123 \\
 51 & A2199    &  445.8 &  20 &   88.2 &  18 &  1298.0 &  14 &    5.1 &  81 \\
 52 & A2204    &   37.6 & 120 &   25.8 &  62 &   674.0 &  29 &    1.5 & 128 \\
 53 & A2244    &   90.7 &  92 &   25.9 &  61 &   400.6 &  44 &    3.5 & 102 \\
 54 & A2256    &  368.1 &  27 &   83.2 &  21 &   841.3 &  22 &    4.4 &  88 \\
 55 & A2255    &  178.9 &  58 &   35.6 &  46 &   253.6 &  66 &    5.0 &  82 \\
 56 & A3667    &  217.6 &  45 &  143.2 &  11 &   780.4 &  26 &    1.5 & 126 \\
 57 & S1101    &   74.0 & 101 &   11.5 &  97 &   339.7 &  52 &    6.4 &  65 \\
 58 & A2589    &  175.6 &  61 &   25.2 &  66 &   283.4 &  59 &    7.0 &  62 \\
 59 & A2597    &   60.7 & 109 &   12.9 &  92 &   427.5 &  38 &    4.7 &  85 \\
 60 & A2634    &  331.5 &  31 &   34.6 &  47 &   177.0 &  79 &    9.6 &  51 \\
 61 & A2657    &  167.9 &  63 &   28.6 &  56 &   246.7 &  68 &    5.9 &  71 \\
 62 & A4038    &  257.0 &  37 &   49.0 &  32 &   592.2 &  30 &    5.2 &  78 \\
 63 & A4059    &  180.9 &  57 &   30.7 &  51 &   427.3 &  39 &    5.9 &  70 \\
\hline
\end{tabular}
\caption{Gamma-ray emission predictions for the groups and clusters in the HIGFLUCS catalog (second part). Fluxes are in units of $10^{-12}{\rm ph}/({\rm cm}^{2}\ {\rm s})$\label{tab:hf2}}
\end{table}

\begin{table}
\begin{tabular}{rlrrrrrrrr}
\hline
ID & Name & DM\_LB & DM\_LB,& CR\_S & CR\_S, & CR\_C & CR\_C,& DM\_LB/CR\_S & DM\_LB/CR\_S, \\
 & & & rank & & rank & & rank & & rank\\
\hline
 64 & A2734    &   90.2 &  93 &   17.0 &  80 &   134.3 &  89 &    5.3 &  77 \\
 65 & A2877    &  440.0 &  21 &   23.3 &  70 &   146.9 &  85 &   18.9 &  28 \\
 66 & NGC499   &  161.2 &  65 &    0.4 & 123 &    26.9 & 110 &  {\color{red}\bf417.1} &   {\color{red}\bf4} \\
 67 & AWM7     & {\color{red}\bf1236.0} &  {\color{red}\bf10} &  {\color{red}\bf149.7} &  {\color{red}\bf10} &  {\color{red}\bf1819.0} &   {\color{red}\bf8} &    8.3 &  57 \\
 68 & PERSEUS  & {\color{red}\bf1940.0} & {\color{red}\bf8} & {\color{red}\bf1353.0} &   {\color{red}\bf1} & {\color{red}\bf20230.0} &   {\color{red}\bf1} &    1.4 & 130 \\
 69 & S405     &  109.4 &  84 &   28.8 &  54 &   137.4 &  87 &    3.8 &  96 \\
 70 & 3C129$^{\dagger}$     & 1126.0 &  11 &  {\color{red}\bf201.2} &   {\color{red}\bf8} &  1189.0 &  18 &    5.6 &  73 \\
 71 & A0539    &  272.3 &  36 &   36.4 &  44 &   261.6 &  63 &    7.5 &  59 \\
 72 & S540     &  138.4 &  75 &   12.1 &  94 &   133.7 &  90 &   11.4 &  43 \\
 73 & A0548w   &   35.7 & 122 &    1.5 & 111 &     8.8 & 117 &   23.9 &  24 \\
 74 & A0548e   &   99.9 &  87 &   16.4 &  83 &    85.7 & 103 &    6.1 &  67 \\
 75 & A3395n   &  377.4 &  24 &   21.8 &  73 &   180.6 &  78 &   17.3 &  30 \\
 76 & UGC03957 &  210.4 &  48 &   10.7 & 100 &   190.9 &  73 &   19.6 &  26 \\
 77 & PKS0745$^{\dagger}$  &   83.0 & 100 &   55.9 &  30 &  1533.0 &  12 &    1.5 & 127 \\
 78 & A0644    &  245.4 &  39 &   44.3 &  37 &   804.5 &  23 &    5.5 &  74 \\
 79 & S636     &  510.1 &  17 &   37.9 &  41 &   165.8 &  82 &   13.5 &  35 \\
 80 & A1413    &   50.5 & 113 &   15.8 &  85 &   249.6 &  67 &    3.2 & 106 \\
 81 & M49      & {\color{red}\bf2009.0} &   {\color{red}\bf6} &    2.8 & 106 &   149.3 &  84 &  {\color{red}\bf710.0} &   {\color{red}\bf3} \\
 82 & A3528n   &   97.2 &  91 &   13.2 &  91 &   113.9 &  98 &    7.3 &  60 \\
 83 & A3528s   &   54.2 & 111 &   22.4 &  72 &   115.8 &  96 &    2.4 & 115 \\
 84 & A3530    &  157.1 &  68 &   12.7 &  93 &    89.4 & 102 &   12.4 &  37 \\
 85 & A3532    &  163.4 &  64 &   25.2 &  65 &   189.0 &  74 &    6.5 &  64 \\
 86 & A1689    &   46.8 & 115 &   15.9 &  84 &   387.7 &  46 &    2.9 & 111 \\
 87 & A3560    &   89.1 &  95 &   18.6 &  77 &    90.6 & 101 &    4.8 &  84 \\
 88 & A1775    &   63.4 & 108 &   14.4 &  88 &   114.8 &  97 &    4.4 &  90 \\
 89 & A1800    &   87.4 &  97 &   14.4 &  87 &   109.8 &  99 &    6.1 &  69 \\
 90 & A1914    &   74.0 & 102 &   16.5 &  82 &   407.3 &  42 &    4.5 &  87 \\
 91 & NGC5813  &  555.6 &  15 &    0.8 & 118 &    54.7 & 107 &  {\color{red}\bf724.7} &   {\color{red}\bf2} \\
 92 & NGC5846  &  853.0 &  12 &    0.9 & 115 &    76.6 & 105 &  {\color{red}\bf943.5} &   {\color{red}\bf1} \\
 93 & A2151w   &  107.6 &  86 &   10.7 & 101 &   126.2 &  93 &   10.1 &  49 \\
 94 & A3627$^{\dagger}$    & {\color{red}\bf2081.0} &   {\color{red}\bf5} &  {\color{red}\bf598.1} &   {\color{red}\bf3} &  {\color{red}\bf3128.0} &   {\color{red}\bf4} &    3.5 & 103 \\
 95 & TRIANGUL &  504.2 &  18 & {\color{red}\bf 233.0} &   {\color{red}\bf7} &  2018.0 &   6 &    2.2 & 120 \\
 96 & OPHIUCHU$^{\dagger}$ & {\color{red}\bf2497.0} &   {\color{red}\bf2} &  {\color{red}\bf459.8} &   {\color{red}\bf4} &  {\color{red}\bf8642.0} &   {\color{red}\bf2} &    5.4 &  76 \\
 97 & ZwCl1742 &  118.5 &  82 &   19.5 &  75 &   261.1 &  64 &    6.1 &  68 \\
 98 & A2319    &  346.5 &  28 &  {\color{red}\bf240.5} &   {\color{red}\bf6} &  {\color{red}\bf1790.0} &   {\color{red}\bf9} &    1.4 & 129 \\
 99 & A3695    &   72.6 & 104 &   29.6 &  52 &   165.0 &  83 &    2.5 & 114 \\
100 & IIZw108  &  125.4 &  80 &   24.0 &  69 &   127.8 &  92 &    5.2 &  79 \\
101 & A3822    &   87.8 &  96 &   28.7 &  55 &   171.1 &  80 &    3.1 & 108 \\
102 & A3827    &  176.9 &  59 &   24.8 &  67 &   320.3 &  54 &    7.1 &  61 \\
103 & A3888    &   99.2 &  89 &   11.5 &  98 &   254.7 &  65 &    8.6 &  55 \\
104 & A3921    &   97.4 &  90 &   15.7 &  86 &   180.8 &  77 &    6.2 &  66 \\
105 & HCG94    &  126.3 &  79 &   18.2 &  79 &   144.0 &  86 &    6.9 &  63 \\
106 & RXJ2344  &  112.2 &  83 &   13.3 &  90 &   181.2 &  76 &    8.4 &  56 \\
\hline
\end{tabular}
\caption{Gamma-ray emission predictions for the groups and clusters in the Extended Sample of the HIGFLUCS catalog. Fluxes are in units of $10^{-12}{\rm ph}/({\rm cm}^{2}\ {\rm s})$. Clusters at low galatic latitudes, $|b|<10$ deg, are marked with $\dagger$. These clusters will have higher gamma-ray backgrounds due to galactic emission.\label{tab:hfext}}
\end{table}

\begin{table}
\begin{tabular}{rlrrrrrrrr}
\hline
ID & Name & DM\_LB & DM\_LB,& CR\_S & CR\_S, & CR\_C & CR\_C,& DM\_LB/CR\_S & DM\_LB/CR\_S, \\
 & & & rank & & rank & & rank & & rank\\
\hline
1 & NGC383   &   57.1 & 110 &    6.5 & 104 &    29.7 & 109 &    8.8 &  53 \\
2 & NGC533   &   36.9 & 121 &    2.0 & 109 &    17.2 & 111 &   18.2 &  29 \\
3 & NGC720   &  150.0 &  71 &    0.6 & 121 &    10.7 & 116 &  {\color{red}\bf267.9} &   {\color{red}\bf7} \\
4 & NGC741   &   42.9 & 117 &    1.1 & 113 &    12.1 & 113 &   39.4 &  20 \\
5 & HCG22    &   17.7 & 126 &    0.05 & 130 &     0.5 & 130 &  {\color{red}\bf383.6} &   {\color{red}\bf5} \\
6 & NGC1407  &  321.0 &  33 &    2.3 & 108 &    93.3 & 100 &  141.5 &  13 \\
7 & NGC1587  &   65.1 & 107 &    0.3 & 124 &     3.9 & 127 &  {\color{red}\bf190.6} &  {\color{red}\bf10} \\
8 & NGC2563  &   33.7 & 123 &    2.3 & 107 &    11.5 & 115 &   14.6 &  34 \\
9 & HCG42    &   40.0 & 118 &    0.2 & 126 &     7.3 & 119 &  {\color{red}\bf206.3} &   {\color{red}\bf9} \\
10 & NGC3557  &   18.0 & 125 &    0.1 & 128 &     5.1 & 123 &  132.9 &  15 \\
11 & NGC3607  &   69.1 & 105 &    1.0 & 114 &     5.9 & 122 &   66.0 &  18 \\
12 & NGC3665  &   49.8 & 114 &    0.1 & 127 &     2.7 & 128 &  {\color{red}\bf352.3} &   {\color{red}\bf6} \\
13 & NGC4065  &   19.6 & 124 &    1.6 & 110 &     7.1 & 120 &   12.4 &  36 \\
14 & NGC4073  &   39.7 & 119 &    9.5 & 103 &    66.5 & 106 &    4.2 &  93 \\
15 & NGC4261  &  176.0 &  60 &   10.2 & 102 &    46.6 & 108 &   17.2 &  31 \\
16 & NGC4325  &   14.3 & 128 &    1.4 & 112 &    14.6 & 112 &   10.6 &  46 \\
17 & NGC4589  &  129.4 &  78 &    0.9 & 116 &    11.6 & 114 &  148.6 &  12 \\
18 & HCG62    &   68.1 & 106 &    5.7 & 105 &   117.2 &  95 &   12.0 &  39 \\
19 & NGC5129  &   12.8 & 129 &    0.9 & 117 &     7.3 & 118 &   15.0 &  32 \\
20 & HCG67    &   10.4 & 130 &    0.1 & 129 &     1.7 & 129 &  170.6 &  11 \\
21 & HCG68    &   54.2 & 112 &    0.6 & 120 &     4.8 & 125 &   94.6 &  17 \\
22 & HCG90    &   44.9 & 116 &    0.7 & 119 &     6.0 & 121 &   60.2 &  19 \\
23 & IC1459   &   73.8 & 103 &    0.3 & 125 &     4.6 & 126 &  {\color{red}\bf239.2} &   {\color{red}\bf8} \\
24 & HCG97    &   17.6 & 127 &    0.5 & 122 &     4.8 & 124 &   36.6 &  21 \\
\hline
\end{tabular}
\caption{Gamma-ray emission predictions for the groups in the GEMS catalog. Fluxes are in units of $10^{-12}{\rm ph}/({\rm cm}^{2}\ {\rm s})$\label{tab:gems}}
\end{table}

\end{document}